%% file: main.tex
\documentclass[conference]{IEEEtran}
\IEEEoverridecommandlockouts

\usepackage{cite}
\usepackage{amsmath,amssymb,amsfonts}
\usepackage[dvipsnames, table]{xcolor}
\PassOptionsToPackage{table,dvipsnames}{xcolor}
\usepackage[hypertexnames=false,hidelinks]{hyperref}
\usepackage{orcidlink}

\usepackage{algorithmic}
\usepackage{graphicx}
\usepackage{textcomp}
\usepackage[normalem]{ulem}
\usepackage{multirow}
\usepackage{booktabs}
\usepackage{fancyvrb}
\usepackage{rotating}
\usepackage{tabularray}
\usepackage{tablefootnote}
\usepackage{printlen}
\usepackage{listings}
\usepackage[]{threeparttable}
\usepackage{listings}

\usepackage{cleveref}
\usepackage{xspace}
\usepackage{paralist}
\usepackage{xargs}
\usepackage{siunitx}

\usepackage{subcaption}
\def\BibTeX{{\rm B\kern-.05em{\sc i\kern-.025em b}\kern-.08em
    T\kern-.1667em\lower.7ex\hbox{E}\kern-.125emX}}

\usepackage{tikz}
\usetikzlibrary{calc}

\usepackage{pgfcalendar}
\newcount\myjuliandate
\newcount\myjuliantoday

\newcommand{\difftoday}[3]{
\textcolor{red}{\makebox[1cm]{
\pgfcalendardatetojulian{\year-\month-\day}{\myjuliantoday}%
\pgfcalendardatetojulian{#1-#2-#3}{\myjuliandate}%
\advance\myjuliandate by-\myjuliantoday\relax
\the\myjuliandate
}}}

\newif\ifdoubleblind
\doubleblindfalse

\usepackage{todonotes}
\newcommandx{\new}[1][]{{#1}}

\definecolor{c1}{RGB}{0,73,114}
\definecolor{c2}{RGB}{238,241,251} 

\newif\ifextended
\extendedfalse

\newcommand{\ready}[1]{#1}

\newcommand{\softname}[1]{\textit{ftio}~}

\ifdoubleblind
\newcommand{\plafrim}{\textit{B}\xspace}
\newcommand{\machine}{\textit{A}\xspace}
\else
\newcommand{\plafrim}{PlaFRIM\xspace}
\newcommand{\machine}{Lichtenberg\xspace}
\fi

\newcommand{\meanshift}{$\phi$\xspace}

\newcommand{\approach}{FTIO\xspace}
\newcommand{\stdtime}{$\sigma_{time}$\xspace}
\newcommand{\stdvol}{$\sigma_{vol}$\xspace}

\newcommand{\halfamplitude}{half-amplitude\xspace}

\definecolor{MidnightBlue}{rgb}{0.0, 0.2, 1}

\crefname{equation}{Eq.}{Eqs.}
\crefrangelabelformat{equation}{(#3#1#4--#5#2#6)}
\crefrangeformat{equation}{#3Eqs. (#1)#4 to #5(#2)#6}
\crefmultiformat{equation}{#2Eqs. (#1)#3}{ and #2(#1)#3}{, #2(#1)#3}{ and #2(#1)#3}
\crefrangemultiformat{equation}{#3Eqs. ((#1))#4 to #5((#2))#6}{ and #3(#1)#4 to #5(#2)#6}{, #3(#1)#4 to #5(#2)#6}{ and #3(#1)#4 to #5(#2)#6}
\crefname{figure}{Figure}{Figures}
\Crefname{figure}{Figure}{Figures}
\crefname{section}{Section}{Sections}
\Crefname{section}{Section}{Sections}

\makeatletter
\newcommand{\linebreakand}{%
  \end{@IEEEauthorhalign}
  \hfill\mbox{}\par
  \mbox{}\hfill\begin{@IEEEauthorhalign}
}
\makeatother
\begin{document}

\title{Capturing Periodic I/O Using Frequency Techniques

\thanks{
This work was funded by the European Commission and the German Federal Ministry of Education and Research (BMBF) under the EuroHPC programmes DEEP-SEA (GA no. 955606, BMBF funding no. 16HPC015) and ADMIRE (GA no. 956748, BMBF funding no. 16HPC006K), which receive support from the European Union’s Horizon 2020 programme and DE, FR, ES, GR, BE, SE, GB, CH (DEEP-SEA) or DE, FR, ES, IT, PL, SE (ADMIRE).
Furthermore, we express our gratitude toward the French National Research Agency (ANR) for the financial support in the frame of DASH (ANR-17- CE25-0004), by the Project Région Nouvelle Aquitaine 2018-1R50119 ''HPC scalable ecosystem''. Moreover, this work received further funding from the German Federal Ministry of Education and Research (BMBF) and the Hessian Ministry of Science and Research, Art and Culture (HMWK) as part of the NHR Program. 
The authors gratefully acknowledge the computing time provided to them on the high-performance computer Lichtenberg at the NHR Center NHR4CES@TUDa. This is funded by the German Federal Ministry of Education and Research and the Hessian Ministry of Science and Research, Art and Culture (HMWK). Computations for this research were performed using computing resources under Project 2215. Finally, the authors also gratefully acknowledge the computing time provided to them on the PlaFRIM experimental testbed supported by INRIA, CNRS (LABRI and IMB), Université de Bordeaux, Bordeaux INP and \emph{Conseil Régional d’Aquitaine} (see \url{https://www.plafrim.fr}).
}

}

\author{
\IEEEauthorblockN{Ahmad Tarraf~\orcidlink{0000-0002-9174-5598}}
\IEEEauthorblockA{
	\textit{Department of Computer Science} \\
	\textit{Technical University of Darmstadt}\\
	Darmstadt, Germany\\
	ahmad.tarraf@tu-darmstadt.de
}
\and
\IEEEauthorblockN{Alexis Bandet~\orcidlink{0009-0000-3822-4374}}
\IEEEauthorblockA{
	\textit{Univ. Bordeaux, CNRS, Bordeaux INP}\\
	\textit{Inria, LaBRI, UMR 5800}\\
	Talence, France\\
	alexis.bandet@inria.fr
}
\and
\IEEEauthorblockN{Francieli Boito~\orcidlink{0000-0002-1139-0724}}
\IEEEauthorblockA{
	\textit{Univ. Bordeaux, CNRS, Bordeaux INP}\\
	\textit{Inria, LaBRI, UMR 5800}\\
	Talence, France\\
	francieli.zanon-boito@u-bordeaux.fr
}

\linebreakand
\IEEEauthorblockN{Guillaume Pallez~\orcidlink{0000-0001-8862-3277}}
\IEEEauthorblockA{
 \textit{Inria}\\
	Rennes, France\\
	guillaume.pallez@inria.fr
}
\and
\IEEEauthorblockN{Felix Wolf~\orcidlink{0000-0001-6595-3599}}
\IEEEauthorblockA{
	\textit{Department of Computer Science} \\
	\textit{Technical University of Darmstadt}\\
	Darmstadt, Germany\\
	felix.wolf@tu-darmstadt.de}
}

\maketitle
\input{00_abstract.tex}


\ifdoubleblind
\else
\fi

\input{01_introduction.tex}

\input{02_approach.tex}

\input{03_evaluation}

\input{04_application}

\input{05_related_work.tex}

\input{06_conclusion.tex}

\input{07_acknowledgement.tex}

\bibliographystyle{IEEEtran}
\bibliography{literature,ref}
\end{document}

%% file: 00_abstract.tex
\begin{abstract}
Many HPC applications perform their I/O in bursts that follow a periodic pattern. 
This allows for making predictions as to when a burst occurs. 
System providers can take advantage of such knowledge to reduce file-system contention by actively scheduling I/O bandwidth. 
The effectiveness of this approach, however, depends on the ability to \emph{detect} and \emph{quantify} the periodicity of I/O patterns online. 
In this paper, we introduce \approach, an online method to detect periodic I/O phases, which is based on discrete Fourier transform (DFT), 
combined with outlier detection. We provide metrics that gauge the confidence in the output and tell 
how far from being periodic the signal is. We validate our approach with large-scale experiments on a production system 
and examine its limitations extensively. Our experiments show that \approach has a mean error below 11\%. Finally, we demonstrate that \approach allowed the I/O scheduler Set-10 to boost system utilization by \new{26\%} and reduce I/O slowdown by \new{56\%}.
\end{abstract}


%% file: 01_introduction.tex
\section{\ready{Introduction}}
\label{sec:introduction}

HPC applications often alternate between compute
phases and access to storage~\cite{carns200924,dorier:hal-01238103,gainaru2015scheduling}.  
Common practices in HPC, such as checkpointing or
visualization~\cite{gainaru2015scheduling,10.1007/978-3-030-29400-7_4}, 
make the I/O phases often periodic, and usually 
involve long file system accesses, which can be a
source for I/O and network contention. Aside from causing performance
variability~\cite{7516071,8752753}, contention means that jobs run longer, harming the platform's utilization and ultimately wasting
resources. 
Solutions proposed to alleviate these aspects include I/O 
scheduling~\cite{gainaru2015scheduling,dorier2014calciom,zhou2015aware,jeannot2021scheduling,boito2023io}, 
I/O-aware batch scheduling~\cite{2626334,3197941,bleuse2018interference,7877148}, and 
the use of burst buffers~\cite{9139729,aupy2018size,aupy2019sizing}.
A challenge when designing such solutions is obtaining 
knowledge of the 
applications' I/O patterns. 

\new{There are several approaches to gathering I/O knowledge depending on the precision needed.}
The most popular tool is probably Darshan~\cite{snyder2016modular}, \new{which gathers aggregated metrics.} However, these aggregated metrics do not properly represent the \emph{temporal} behavior of applications~\cite{9820616}. Because I/O tends to be bursty and periodic, knowing how many bytes 
are accessed does not paint the full picture. We need to know \emph{when} (or rather \emph{how often}) these accesses happen: two applications that write each 1~TB over 2 hours, one with a single I/O phase at the end of the execution and the other 
one with multiple I/O phases every 2 minutes, impose very 
different loads on the system.
\new{Thus, one might desire a detailed description of I/O activity over time. However, finding an extremely precise profile comes at the cost of a higher overhead, both in terms of measurement and data accumulation. Moreover, a detailed time model can be hard to 
explore in a contention-avoidance algorithm that is lightweight enough to be used in practice, 
%
especially as models with high predictive accuracy are
often \emph{black box} and cannot be interpreted directly for explaining I/O performance~\cite{9355272}. 
Hence, depending on the use case, models at higher abstraction levels might be tolerated, which can be easier to interpret and often to generate, especially online. 
}

\new{Recent work} on I/O scheduling~\cite{boito2023io,benoit:hal-04038011,jeannot2021scheduling,dorier2014calciom} 
has shown that knowledge of periodic I/O patterns, even when not \emph{perfectly precise}, leads to good contention avoidance.
\new{Consequently, one approach could be to predict the \emph{period of the I/O phases} during runtime and provide this information to such approaches rather than finding detailed time models.}
\new{This is the metric we seek in this work, which presents a trade-off between aggregated information and a detailed time model.}
 
\begin{figure}[thbp]
    \centering
    \includegraphics[width=.95\linewidth]{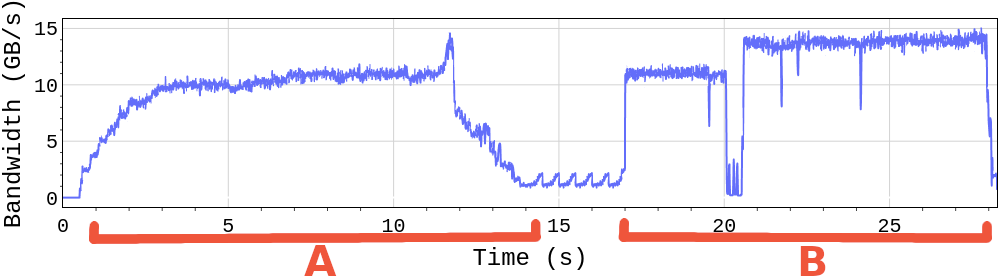}
    \caption{\new{Difficulty of detecting I/O phases: Where does A finish? Is B one or two phases? Why don't A and B belong together?}}
    \label{fig:exampleIO}
\end{figure}

%
%
%
\new{However, describing} the temporal I/O behavior in terms of I/O phases is a challenging task.
Indeed, the HPC I/O stack only sees a stream of issued requests and
does not provide I/O behavior characterization.
Contrary, the notion of an
\emph{I/O phase} is often purely logical, as it may consist of a set of
independent I/O requests, 
issued by one or more processes and threads during a particular time window, and popular 
APIs do not require that applications explicitly group them. 
Thus, a major challenge is to draw the borders of an \emph{I/O phase} (see \cref{fig:exampleIO}).
%
Consider, for example, an application with 10 processes that writes $10$~GB by generating a sequence of two $512$~MB write requests per process,
then performs computation and communication for a certain amount of time, after which it writes 
again $10$~GB. How do we assert that the first $20$~requests
correspond to the first I/O phase and the last $20$ to a second one?
An intuitive approach is to compare the time between consecutive requests with a
given threshold to determine whether they belong to the same phase. Naturally,
the suitable threshold should depend on the system. The reading or
writing method can make this an even more complex challenge, as accesses can
occur, e.g., during computational phases in the absence of barriers.
Hence, the threshold would not only be \emph{system dependent} but also
\emph{application dependent}, making this intuitive approach more complicated
than initially expected.

Even assuming that one is able to find the boundaries of various I/O phases,
this might still not be enough.
Consider, for example, an application that periodically writes large checkpoints with all processes. In addition, a single process writes, at a different frequency, only a few bytes to a small log file.
Although both activities clearly constitute I/O,
only the period of the checkpoints is relevant to
contention-avoidance techniques. If we simply see I/O activity as belonging to
I/O phases,
we may observe a
profile that does not reflect the behavior of interest very well. 

Thus, a method for characterizing the temporal I/O behavior of an application is
needed \new{that determines the period of the I/O phases and is thus at a higher abstraction level compared to detailed time modeling approaches.} 
To be useful in practice, it should impose minimal overhead and generate only a modest amount of
information, especially during the \emph{online execution} of an application.
This is where our paper aims to contribute:
\begin{itemize}
    \item We propose \approach, which characterizes the temporal I/O behavior of an
application in terms of its period, obtained using frequency techniques.
Additionally, we provide strategies to adapt to behavioral changes.
	
    \item We introduce metrics that quantify the confidence in the obtained
results and further characterize the I/O behavior based on the identified period. 

    \item \approach is implemented as an open-source library, which can be easily
    attached to existing codes, 
    to provide online predictions of their I/O behavior with low overhead. Moreover, we offer an offline realization as well.

    \item We evaluate \approach with large-scale applications and extensively 
    study its limitations and accuracy using traces crafted to represent challenging situations. Moreover, we show how \approach enables an I/O scheduler to reduce I/O slowdown by \new{56\%} and boost system utilization by \new{26\%}.

\end{itemize}

By finding the period of I/O, the average amount of data and time spent per I/O phase can be calculated, 
which has clear usefulness for burst buffer management, for example. 



This paper is organized as follows: 
In Section~\ref{sec:ftio}, we present our approach: 
how we collect 
information, DFT and how we use it,
the additional confidence metrics, 
and the implementation of \approach.
Our strategy is evaluated 
in Section~\ref{sec:evaluation},
while in Section~\ref{sec:motivation} we illustrate the use of FTIO for I/O scheduling. 
Finally, we discuss related work in Section~\ref{sec:related}, before concluding
on the implications of this work in Section~\ref{sec:conclusion}.

%% file: 02_approach.tex
\section{\approach: Finding the Period of I/O}
\label{sec:ftio}

This section presents our approach 
to characterize the temporal I/O behavior of an HPC application in terms of the period of its I/O phases.
We call our methodology: Frequency Techniques for I/O (\approach), as it leverages well-known signal-processing techniques.
\approach is implemented as a two-step approach: 
\begin{inparaenum}[(1)]
\item a library at the application side intercepts the I/O calls
and appends the collected data continuously to a file (\cref{subsec:gathering_io}), 
which 
\item can be evaluated at any time (Sections~\ref{sec:prediction} till 
\ref{sec:param}),
on a cluster or a local machine, to determine the period of the I/O phases.
\end{inparaenum}
Both the tracing library named TMIO\footnote{\url{https://github.com/tuda-parallel/TMIO/}\label{footnote:tmio}} (Tracing MPI-IO) and \approach\footnote{\url{https://github.com/tuda-parallel/FTIO/}} are publicly available on GitHub. In what follows, we describe TMIO as part of FTIO.


\subsection{\ready{Gathering the I/O Information}}
\label{subsec:gathering_io}

As the first step of \approach, the I/O information from the application needs
to be collected. 
For this, we developed a tracing library in C++ 
(TMIO) 
that intercepts specific MPI-IO calls to gather metrics such as start time, end time, and transferred bytes. 
We provide two methods for linking the library to the application, 
depending on whether the information is used for offline (\textit{detection}) or 
online (\textit{prediction}) periodicity analysis. 
The offline mode uses the LD\_PRELOAD mechanism. Upon MPI\_Finalize, the collected data is written to a single file to be analyzed later. 
In the online mode, the application is compiled with our library and a single line is added to indicate when to flush the results out to a file (JSON Lines or MessagePack\cite{msgpack}). 
This file can be evaluated anytime using a Python script to dynamically predict the period of the next phases based on the data collected up to this point. 
Note that at the end of the run, the same file can be used for offline evaluation. 
Our library has low overhead as the I/O data is collected at the rank level (individual requests).  
The overlapping of the requests (i.e., bandwidth at the application level) is evaluated by the Python 
script (either entirely or for a given time window) with a linear complexity with the number of I/O requests. 

What we need for the next step 
is essentially the variation of the bandwidth over time.
As the analysis is at the application level, we merge the information we collected per process.
Note that although we used our library in this paper, 
it could easily be replaced by other tools and data sources (e.g., file system monitoring data if available) 
For the \emph{detection} approach, for example, we support Recorder~\cite{CW20} 
and Darshan~\cite{carns200924,snyder2016modular} profile and traces.
The next section describes how the collected information $x(t)$ (bandwidth over time) is further processed.

\subsection{Extracting the Period of I/O}
\label{sec:prediction}
\label{subsec:DFT}
\label{subsec:anomaly}


%

With our approach, we move away from detailed modeling and focus on a
simple metric: The period of the I/O phases. For that, we examine the I/O behavior in the
frequency domain instead of traditionally analyzing it in the time domain. 
Thus, we treat the I/O bandwidth over time as a \emph{signal}, which we first discretize and
then analyze using the discrete Fourier transform (DFT).
\new{
Since we aim to find the period of a signal rather than fully model its time behavior, frequency analysis coupled with outlier detection perfectly suits this task. Compared to time analysis, frequency techniques, such as DFT, decompose a signal into its frequency components, giving us control over the \emph{interesting} I/O. 
Moreover, as we focus on I/O phases rather than individual requests, by applying DFT on the application-level signal, we overcome the challenges from \cref{sec:introduction}. 
Additionally, the parameters of DFT allow \approach to adapt to changing I/O behavior and specify the range of interesting I/O as 
handled later in Sections~\ref{subsec:online} and \ref{sec:param}.
}


\subsubsection{DFT} DFT decomposes a signal into its frequency components 
such that their sum allows reconstructing the signal. 
As an input of DFT, the continuous signal $x(t)$ is discretized 
with a sampling frequency $f_s$ 
to obtain $N = \Delta t \cdot f_s$
samples: 
\begin{equation*}
\{x_n = x(n / f_s)\,|\, n \in [0, N)\}
\end{equation*}
DFT transforms this evenly-spaced sequence from the time domain into a
sequence $X_k$ of the same size with $k \in [0, N)$ bins in the frequency domain: 
\begin{equation*}
X_k = \sum_{n=0}^{N-1}x_n e^{\frac{-2\pi k n}{N}i},
\end{equation*}

\noindent for the frequencies: 
\begin{equation*}
f_k = \frac{k}{N}f_s = \frac{k}{\Delta t}  
\end{equation*}
Thus, DFT is evaluated for the fundamental frequency $\frac{f_s}{N}$ and its harmonics.
As consecutive frequencies $f_k$ are spaced $1/\Delta t$ apart, the
larger the time window $\Delta t$, the closer the components $X_k$ are to each other,   
thus improving the precision of DFT.
However, this increases the complexity of the analysis. 
\ifextended
\else

Since the sampled signal $x_n$ consists of purely real values for our purposes (I/O signal), DFT is symmetric and only
half of the frequencies are needed to reconstruct the original signal 
with the inverse DFT (IDFT):
\begin{equation}
     x_n   = \frac{1}{N} \left (X_0 + \sum_{k=1}^{\frac{N}{2}}2|X_k|\cos{ \left(\frac{2\pi k n}{N} + arg(X_k) \right )} \right )
     \label{eq:simplefied_dft}
\end{equation}
with the amplitude $|X_k|$ and the phase $arg(X_k)$. 
This reduces the calculation needed for the reconstruction of the signal and 
limits the constituting signals to cosine waves only, simplifying the
interpretation of results. 
Consequently, when plotting the amplitude $|X_k|$ against
the frequencies $f_k$, only half of the spectrum
(\emph{single-sided spectrum}) needs to be inspected. In this case, the
amplitude of the fundamental and harmonics need to be multiplied by
two, as shown in \cref{eq:simplefied_dft}.

$X_0$, the DC offset in signal analysis terminology, is expected to be among the highest components as the I/O data 
transferred is always a positive number of bytes, and the cosine waves obtained
with DFT need to be shifted upwards. 
\fi
%
%
Note that, to compute DFT, we use the Fast Fourier Transform (FFT)
algorithm, which has a complexity of $O(NlogN)$.

After obtaining the amplitude spectrum from DFT, we need to examine it to extract 
the period of I/O phases. 
However, I/O tends to exhibit variations and is often affected by noise, which results in high frequencies with small amplitudes. To
alleviate this effect, we use the \emph{power spectrum} ($p_k=\frac{1}{N}X_k^2$) 
instead of the amplitude spectrum. 
By normalizing the power spectrum over the total power of the signal for the
plots, the y-axis 
of the normalized power spectrum 
indicates the \emph{contribution} of the
frequency to the total signal power. 

\subsubsection{Outlier detection}
The most straightforward approach for extracting the period of the I/O phases is to
find the frequency with the highest contribution, i.e., the \emph{dominant
frequency} $f_d$, while excluding the DC offset from the analysis. 
Intuitively, the period is simply then $1/f_d$.
However, if there are multiple frequencies with similarly high contributions, the
frequency with the maximum contribution does \emph{not} properly represent the
temporal behavior. Notably, that is the case for non-periodic signals. Hence,
simply selecting the maximum would silently accept a result that is probably
inaccurate; it also has to be an {\em outlier}. 
%
%
%
One 
approach for detecting outliers is the
Z-score~\cite{kannan2015labeling}. It reveals how many standard deviations
$\sigma$ a power $p_k$ is from the mean $\bar{p}$ of all powers:
\begin{equation}
    z_k =\frac{|p_k| - |\bar{p}|}{\sigma}
    \label{eq:zscore}
\end{equation}
For each frequency $f_k$ with $k\in[1,\frac{N}{2}]$, a Z-score $z_k$ is found.
A Z-score beyond 3 usually indicates an outlier~\cite{kannan2015labeling}.
As there might be several outliers, to find the dominant frequencies, we 
compare the Z-scores of the outliers to 
the largest Z-score
$z_{\text{max}}=\max_{k\geq 1}(z_k)$. 
Moreover, if a frequency $f_k$ is an outlier ($z_k>3$), and its Z-score is
within 80\% of the largest Z-score (a tolerance value that can be adjusted)
($z_k / z_{\text{max}} \geq 0.8$), then it belongs to the set of dominant
frequency candidates $\mathcal{D}_f$:
\begin{equation}
\mathcal{D}_f = \{f_k \;|\; z_k \geq 3 \text{ and } z_k / z_{\text{max}} \geq 0.8\}
    \label{eq:maxzk}
\end{equation}
Depending on the number of candidates, we distinguish several cases.
If $\mathcal{D}_f = \{f_k\}$ (single candidate frequency), we have a high
confidence that the signal is periodic with the dominant frequency $f_d = f_k$.
If $\mathcal{D}_f = \{f_{k_1},f_{k_2}\}$ (two candidate frequencies), the
signal has some variation in its behavior but is still periodic. In this case,
\approach returns that the dominant frequency is the one with the highest power contribution.
Finally, if none or more than two candidates were found, there is no
dominant frequency, and the signal is most likely not periodic.
There is an exception when the candidates are multiples of two of each other.  
In this case, the higher frequencies are ignored. 
The presence of this kind of harmonics with decreasing high contributions 
is an indication that there are periodic I/O bursts in the signal.

We favored simple calculations in our approach, as we aimed for an implementation with
minimal overhead. 
Aside from the Z-score, \approach supports other 
outlier detection 
methods, including DBSCAN, isolation forest, local outlier factor, and the find\_peaks algorithm from SciPy that can all deliver decision functions to find the outliers. 
A key advantage is that several parameters these algorithms require 
can be easily found. For DBSCAN, for example, the frequency step can be used to compute eps (minimal distance between two points to be considered as neighbors). Still, while these algorithms can 
improve the results (either alone or by merging their result with the Z-score), they often require more computational effort. 
Thus, the decision to use them depends on the intended use of \approach.
%
In the next section, we provide metrics that express the confidence in the results of the period extraction.

\subsection{Confidence Metrics}
\label{subsec:autocorrelation}
\label{subsec:additional_characterization}
For a better interpretation of the results, FTIO provides the confidence
$c_k$ in the frequency $f_k$ 
in case at most two frequencies are in $\mathcal{D}_f$. 
If we call $\mathcal{I}_1=\{i \;| \;z_i \geq 3\}$ the set of frequencies that are outliers, and $\mathcal{I}_2=\{i \;| \;z_i/z_{\text{max}} \geq 0.8\}$ the set of frequencies whose Z-score is within 80\% of the maximum Z-score, then: 
\begin{equation*}
    c_k = \frac{1}{2} \left (\frac{z_k}{\sum_{i \in \mathcal{I}_1} z_i} + \frac{z_k}{\sum_{i \in \mathcal{I}_2} z_i}\right )
    \label{eq:conf}
\end{equation*}
Consequently, $c_d$ is the confidence of the dominant frequency $f_d$. 
To refine this confidence metric,
we optionally provide a second 
method that does not rely on the result from DFT, namely autocorrelation. 

\paragraph*{Autocorrelation}
Another 
signal analysis method for finding the period is autocorrelation~\cite{Nounou_Bakshi_2000}. 
The autocorrelation function (ACF) measures the correlation of the observations within a time
series at various lags~\cite{Box_Jenkins_1976}. This allows for spotting repeated patterns in a signal. 
The ACF can attain values in $[-1,1]$. 
We compute the ACF using NumPy's correlate function on the discretized signal with all $N$ samples.
To find the periods in the signal, 
we detect the peak locations in the ACF, find the number of samples \emph{between} them, and divide the obtained values by $f_s$. Unlike DFT, these candidates can be repeated several times. Hence, after filtering outliers (e.g., with Z-score), we find the period of the signal using the average. Using the coefficient of variation ($\sigma/\bar{p}$) we provide a confidence ($c_a=1-\sigma/\bar{p}$) in our result from autocorrelation.

As the period from the autocorrelation is found using averaging, we trust the DFT result more. 
Consequently, if the results are merged, the autocorrelation results are used to adjust the confidence obtained from DFT. 
For that, we find the similarity of the dominant frequency from DFT to the candidates 
from the autocorrelation using the coefficient of variation $c_s$. 
Finally, the \emph{refined confidence} is computed by averaging $(c_d + c_a + c_s)/3$.
Thus, the \emph{refined confidence} is more reliable as different methods found a similar solution.

\begin{figure}[bp]
	\includegraphics[width=.95\columnwidth]{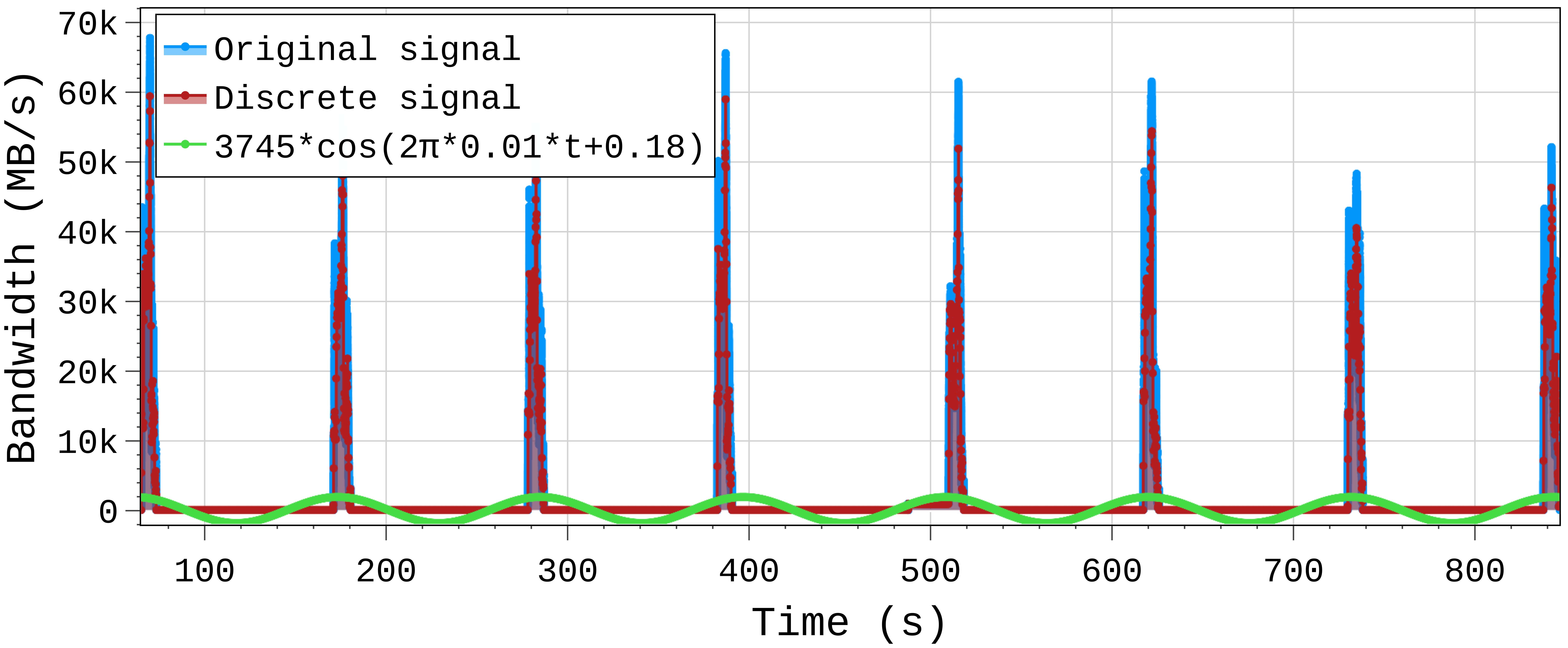}\\[0.3cm]
\includegraphics[width=\columnwidth,trim= 0 0 0cm 0cm, clip]{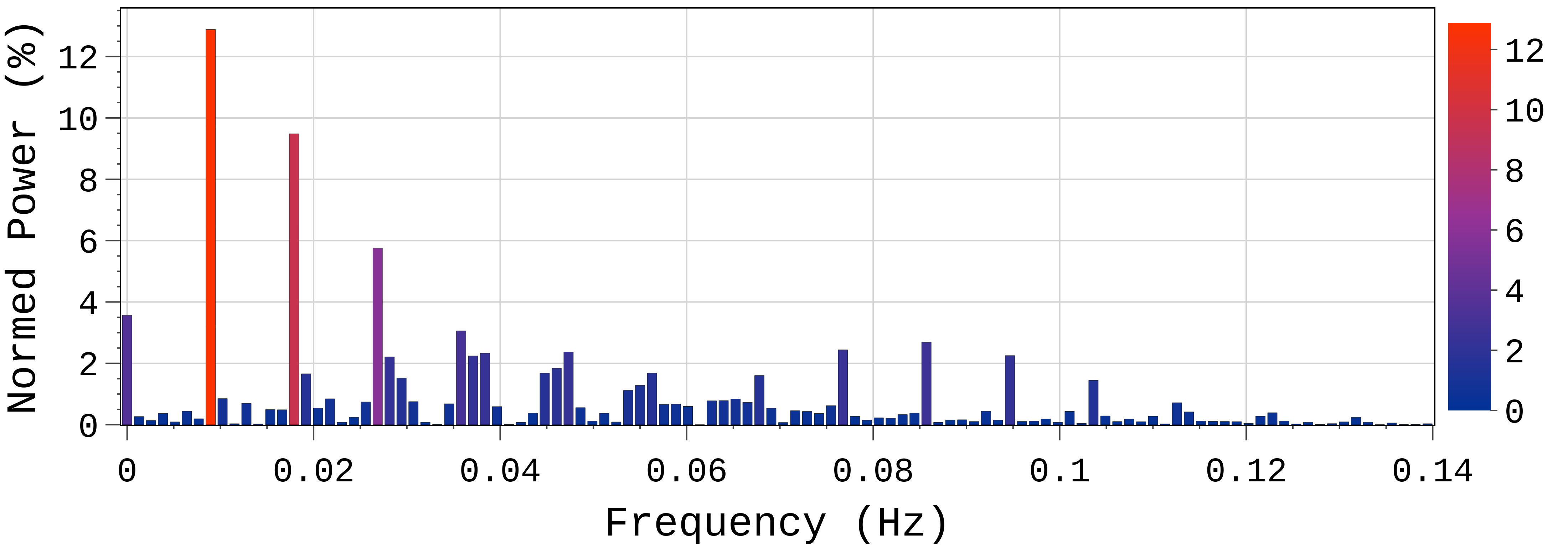}
	\caption{\new{\approach results on IOR with 9216 ranks executed on the Lichtenberg cluster. The time behavior (top) and the 
	normed power spectrum (bottom) are shown.}} 
	\label{fig:ior_9216}
\end{figure}

\paragraph*{Practical example}
We executed
the IOR benchmark with 9216 ranks on \machine cluster (described in
\cref{sec:results}). 
We set up IOR with 8 iterations, 2 segments, a transfer size of 2~MB, 
and a block size of 10~MB with the MPI-IO API in the parallel mode and our library preloaded.
After the execution on the cluster, we run \approach on the result for the entire 
time window $\Delta t$ of 781~s (i.e., from 64.97~s  to 846.7~s) with a sampling frequency $f_s=10$~Hz. 
This resulted in 7817 samples and an abstraction error (difference between discrete and original signal) of 0.03. 
As we only examine half the power spectrum, the number of inspected frequencies is 3809 and the maximum value on the x-axis of the spectrum is 5~Hz.
\approach detected that the signal has a period of 111.67~s (i.e., 0.01~Hz) as 
the top part of \cref{fig:ior_9216} (cosine wave frequency) shows, with a confidence of  $c_d = 60.5\%$. 
The lower part of \cref{fig:ior_9216} shows the normed power spectrum zoomed to the relevant frequencies. 
%
%
%
%
On average, each of the 3809 frequencies contributed 0.025\% to the power. 
As shown, the frequency at 0.01~Hz 
has the highest contribution. If the tolerance value is lowered from 0.8 to 0.45, the frequency at 
0.02~Hz becomes a candidate as well. However, it is a harmonic (multiple of 2) of the dominant frequency and hence is ignored. This increases the confidence to $c_d = 62.5\%$, as the number of candidates that satisfy $z_k \geq 3$ decreases. 
As observed in the figure, the presence of such candidates for the dominant frequency indicates the presence of periodic I/O \emph{bursts} in the signal.

\begin{figure}[htp]
\includegraphics[width=\columnwidth,trim= 0 0 0.2cm 0cm, clip]{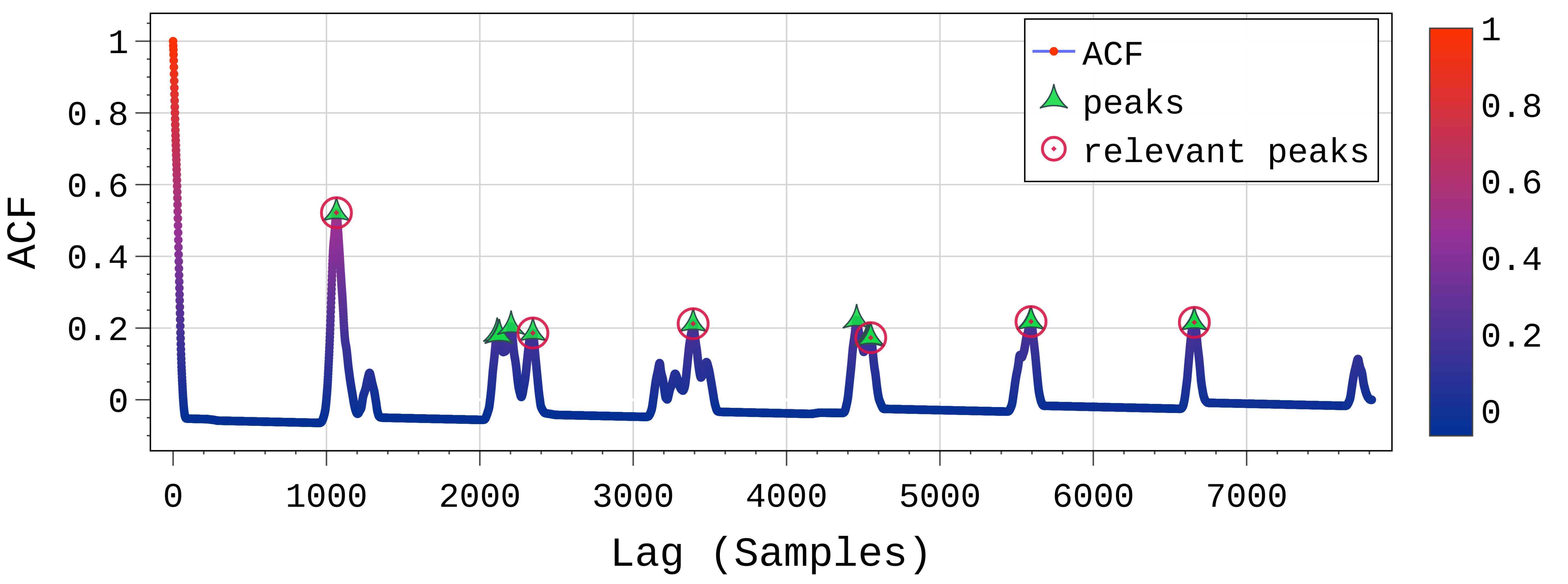}
	\caption{\new{Result of autocorrelation on IOR with 9216 ranks.}} 
	\label{fig:ior_9216_auto}
\end{figure}

In \cref{fig:ior_9216_auto}, the ACF is plotted against the lag measured in samples for the same signal. 
Clearly, the correlation of the signal with itself at zero lag is one. Using the find\_peaks algorithm from SciPy (with a threshold of 0.15), we detected the peaks in the ACF (marked as green triangles in the figure). 
%
%
%
%
Next, we divide the samples between consecutive peaks 
by $f_s$ to obtain 17 periods. Using the Z-score with the \emph{weighted mean} (weights from the ACF), we filter the 12 outliers and thus find 5 candidates.
Finally, we compute the average with the candidates to obtain a period of 104.8~s (i.e., 0.01~Hz). Note, that these candidates are found using the number of samples \emph{between} the peaks which are marked with red circles in \cref{fig:ior_9216_auto}. 
Using the coefficient of variation, we obtain a confidence $c_a=99.58\%$ in the result from autocorrelation. 
Finally, we compute the similarity of $f_d$ 
from DFT to the 5 candidates from the autocorrelation ($c_s = 97.6\%$),   
and average the three values (62.5\%, 99.58\%, and 97.6\%) to obtain a refined confidence of 86.5\%. 
Thus, by additionally using autocorrelation, we can refine the confidence in the results.

\paragraph*{Further characterization}
Often, we want to know how much the I/O phases match the result of \approach or  
how far from being periodic a signal is, or how long an I/O phase is inside a period.
In the rest of this section, we provide metrics to support these aspects and thus allow for additional characterization of the result from \approach. 

\paragraph{Standard deviation of volume (\stdvol)} 
If we assume an application is periodic, and we know its frequency, then in every period roughly the same amount of data is transferred.
For an I/O trace $\mathcal{T}$, let $V(\mathcal{T})$ be the amount of data accessed in it (i.e., the \emph{volume} of I/O).
Given $f_d$ from \approach, we divide the trace into 
sub-traces $\{\mathcal{T}_1,\cdots,\mathcal{T}_{m}\}$ each of length $1/f_d$ and data volume V($\mathcal{T}_i$) for 
$i \leq m$.
We compute \stdvol as the standard deviation of $\frac{V(\mathcal{T}_i)}{\max (V(\mathcal{T}_i))}$. 
The lower this value, the more similar the data volumes accessed per period are, and thus the more periodic the signal. 

\paragraph{Time ratio spent on substantial I/O ($R_{IO}$)}
An application could be periodic (in time) but not accessing the same amount of data per I/O phase (i.e.,  high \stdvol).
We define \stdtime 
to evaluate the (time) periodic behavior. Before that, however, we first need to
define the time ratio spent on substantial I/O.

\begin{figure}[htbp] 
    \centering
    \includegraphics[width=\linewidth]{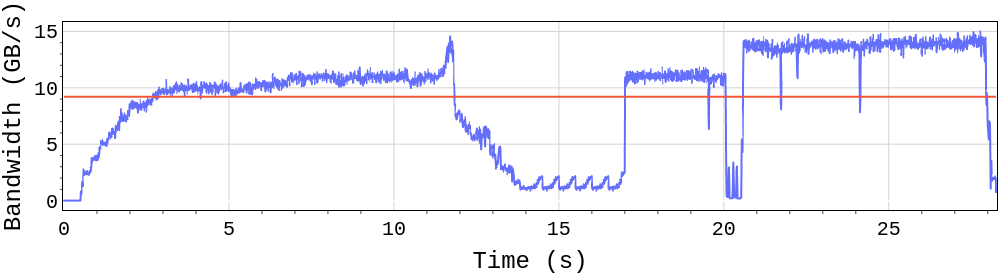}
    \caption{\new{A red line marks the 
      $V(\mathcal{T})/L(\mathcal{T})$ threshold in the trace from Figure~\ref{fig:exampleIO}. Here $R_{IO} = 0.68$ and $B_{IO}\approx11$ GB/s.}}
    \label{fig:bio_and_rio}
\end{figure}

Consider, for example, an application that has frequent low-bandwidth I/O, constantly writing a small log file, interleaved with periodic higher-bandwidth I/O phases. 
In this case, we consider the low-bandwidth activity as {\em noise} and the I/O
phases as \emph{substantial} I/O.  
On the contrary, for a signal composed only of the same low-bandwidth ``noise,''
we might not want to consider it as noise but as the I/O behavior of the 
application. Therefore, we need a threshold of what is noise and what is
not for each application. A fixed per-system threshold could
be enough for some usage of this method (e.g., I/O scheduling), but
here, we focus on the more challenging and generic case.
For the trace $\mathcal{T}$ of length $L(\mathcal{T})$, 
we set the threshold as $V(\mathcal{T})/L(\mathcal{T})$. 
Let $\mathcal{S}$ be the subset of 
the trace where the volume of I/O per time-unit is greater than this threshold.
Having filtered out the noise, we can compute the time ratio spent doing substantial I/O: 
\begin{equation*}
R_{IO} = L(\mathcal{S})/L(\mathcal{T}),  
\end{equation*}
with $R_{IO}\in[0,1]$.
We can also identify the 
        bandwidth characterizing the {\em substantial} I/O of the whole trace: 
\begin{equation*}
B_{IO} = V(\mathcal{S}) / L(\mathcal{S})
\end{equation*}
This is illustrated in \cref{fig:bio_and_rio}.
Moreover, the amount of data transferred per period can be easily calculated by $\frac{V(\mathcal{S})}{L(\mathcal{T})\times f_d}$. The lower \stdvol, the better this value works as a prediction for a future I/O phase.

\paragraph{Standard deviation of time (\stdtime)}

Similarly to
$\mathcal{S}$, let $\mathcal{S}_i$ be the subset of $\mathcal{T}_i$ where the
volume of I/O per time-unit is greater than $V(\mathcal{T})/L(\mathcal{T})$.
%
Then \stdtime is defined as: 
\begin{equation}
    \sigma_{time} = \sqrt{\frac{1}{L(\mathcal{T})\cdot f_d} \sum_{i=1}^{L(\mathcal{T})\cdot f_d}\left (\frac{L(\mathcal{S}_i)}{L(\mathcal{T}_i)} - R_{IO} \right )^2}
\end{equation}

Thus, \stdtime is the standard deviation of the proportion of time spent on I/O \emph{inside each period}. 
The intuition is that, if the signal is periodic and the application spends, e.g., $60\%$ of its time on I/O ($R_{IO}=0.6$), then each of its I/O phases 
will last approximately $60\%$ of a period. Therefore, the lower the \stdtime, the more periodic the signal is expected to be. 

Values close to zero for both \stdtime and \stdvol indicate a signal that is periodic and, therefore, additionally increase our confidence in the period obtained with \approach. On the other hand, a high \stdvol with a low \stdtime indicates the application is probably periodic but does not access similar amounts of data per I/O phase. 
Since both \stdvol and \stdtime are in $[0,0.5]$,
we can provide a \emph{periodicity score} (in $[0,1]$) for the signal according to the \approach-provided period as $1 - $\stdvol$ - $\stdtime.




\subsection{\ready{Online Period Prediction}}
\label{subsec:online}
\begin{figure}[thp]
	\centering
	\includegraphics[width=.9\columnwidth,trim= 0 0 0 0, clip]{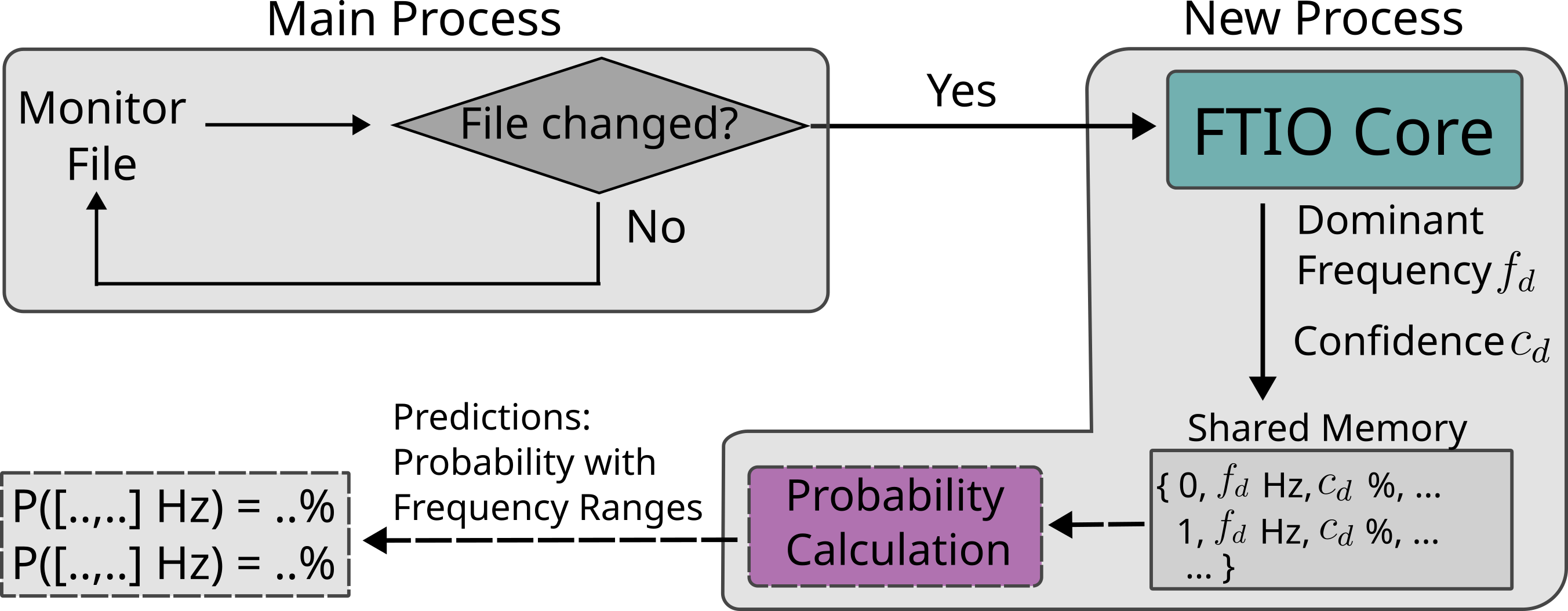}
	\caption{Online period prediction using \approach.} 
	\label{fig:ftio_online}
\end{figure}

So far, we described the offline (post-mortem) detection approach. For online prediction (during application execution),
the approach is similar, with the difference that \approach is executed, to find $f_d$ and $c_d$ (if any), in a new child process every time new I/O measurements are appended to the trace file. \cref{fig:ftio_online} shows an overview of the 
online methodology. 
To adapt to changing I/O behavior and variability, our algorithm 
offers two optional enhancements: (1) adapting time windows and (2) probability calculations with frequency intervals. 

As the I/O behavior of an application can change, it makes sense to discard 
the old data at some point, and hence, consider a shorter time window for the analysis. Different strategies 
can be used here. 
The simplest one is that after finding $k$ times a dominant frequency, the time window for evaluation is reduced to $k$ times the last found period. 
Alternatively, one could specify a fixed length or a fixed $k$. Time window adaptation is demonstrated later in \cref{sec:results}. 

The second enhancement uses the results from consecutive \approach evaluations, 
which are stored in a shared memory between the processes.
As different executions usually have different time windows, the resolution in the frequency domain changes (see \cref{subsec:DFT}). Consequently, when merging predictions, intervals are used. For that, our approach merges the dominant 
frequencies using DBSCAN with eps set to the difference between the time windows. For each cluster, an interval is calculated by finding the minimum and maximum of the dominant frequencies contained in the cluster.
Moreover, the number of predictions inside a cluster 
divided by the total number of predictions 
represents the probability of the interval. 




\subsection{\ready{Parameter Selection}}
\label{sec:param}
Three parameters affect the analysis: 
the time window $\Delta t$, the sampling frequency $f_s$, 
and the number of samples $N=\Delta t \cdot f_s$.
%
The granularity at which the data is captured is specified with $f_s$. 
As our approach captures the time spent on each I/O request, 
we can find the smallest change in bandwidth over time and use it to calculate $f_s$.
However, this is often unnecessary, as we are usually not interested in high-frequency behavior.
In contrast, a too-low sampling frequency could result in aliasing. 
\ifextended
\else
The importance of this is illustrated in  \cref{fig:miniio}, which shows the results of 
\approach on miniIO~\cite{minio} executed with 144 ranks on the \machine cluster. 
%
%
\begin{figure}[thp]
    \centering
    \includegraphics[width=\columnwidth]{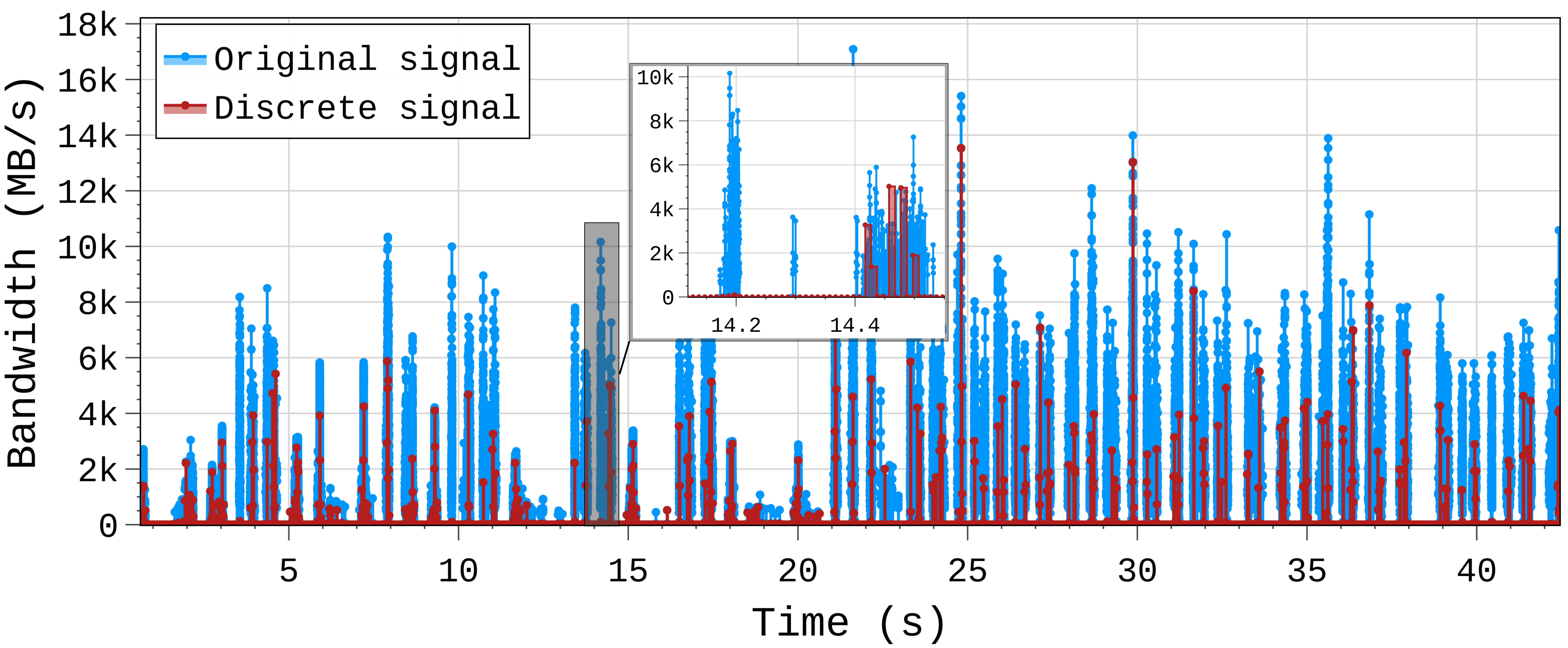}
    \caption{\new{miniIO with 144 ranks on the \machine cluster.}} 
    \label{fig:miniio}
\end{figure}
The \textit{unstruct} mini-app was used, which produces unstructured grids with 1000 points per task. 
In \cref{fig:miniio}, we set $f_s$ to 100~Hz, which is not enough: the discrete signal does not match 
the original one at all.  
But even if the approach had found a single period, the result cannot be trusted, as the abstraction error 
(the volume difference between the two shown signals) is just too large. 
\fi


With a constant sampling frequency $f_s$, increasing $\Delta t$ increases the number of samples  $N=\Delta t \cdot f_s$, which increases 
the detection/prediction time. 
In all of our experiments, this time was negligible. 
Moreover, it does \emph{not} represent overhead to applications, since the
analysis is \emph{not} done on the nodes where they run. The only overhead there comes from 
the tracing library and is analyzed in \cref{subsec:overhead}.

%% file: 03_evaluation.tex
\section{Evaluation}
\label{sec:evaluation}

In this section, we evaluate \approach by: (1) analyzing its accuracy and limitations (\cref{sec:results_limits}), 
(2) demonstrating it based on three case studies  (\cref{sec:results_validation}), and (3) 
examining its overhead  (\cref{subsec:overhead}). 

\input{03_00_limits}
\input{03_02_results_validation}

\input{03_01_overhead}

%% file: 03_00_limits.tex
\subsection{Limitations of FTIO}
\label{sec:results_limits}
In what follows, we explore the accuracy and limitations of \approach by crafting challenging synthetic traces. 

\input{03_00_01_limits_methodology}

\paragraph*{Results}
\input{03_00_02_error_figures}
First, we study
the impact of length differences between CPU and I/O phases, e.g., as seen in 
\cref{fig:ior_9216} and later in \cref{lammps}.
For this, we use traces with $\delta_k=0$ and vary $t_{cpu}$ (with $\sigma=0$).
\cref{fig:limit_meancpu} presents the results and shows that the disparity in phase duration is \emph{not} 
a problem. They also seem to indicate that when the time between I/O phases is \emph{longer}, 
our approach leads to better results. However, that might be an artifact of the fixed sampling frequency. 
Still, all errors are below $1\%$. These results also suggest that \approach is fairly 
robust to noise.

Next, we cover two challenging scenarios at once: 
\begin{inparaenum}[(1)]
	\item when the processes performing the I/O phase are not synchronized (absence of implicit/explicit barriers), and
	\item I/O performance variability, with some I/O phases of the application taking longer than others, which is usually the case when accessing a shared file system.	
\end{inparaenum}
For that, we set $t_{cpu}=11$~s and increase \meanshift (the average $\delta_k$). 
The results are shown in \cref{fig:limit_shift}. 
When \meanshift becomes larger than the original duration of I/O phases, there are often periods without 
I/O activity inside the I/O phases, which makes their detection more difficult. In extreme cases, 
the error goes up to 100\%, but is in general low: Mean of up to 11\%, median up to 11\%, and third quartile up to 17\%. 

Finally, we study the case where the time between I/O phases varies during the execution,  
as in \cref{fig:ior_9216,lammps}. 
We control that by drawing $t_{cpu}$ from $\mathcal{N}(\mu,\sigma)$ with $\mu=11$~s 
and increasing $\sigma$. For this experiment, we use $\delta_k=0$ and no noise. Note that
the use of real I/O traces for the phases introduces natural variability. 
In \cref{fig:limit_cpusigma}, we can see the results vary in quality as the signal becomes less periodic. 
This figure was zoomed-in to allow the visualization of the box plots. 
$26$ outliers with errors of more than 200\% are not shown (out of 3400 traces). 
They are: 0.4\% of the traces with  
$\mu \leq \sigma < 2\mu$, and 1.9\% of the traces with $\sigma \geq 2\mu$. 
With $0.5\mu \leq \sigma < \mu$,
16\% of the traces obtained confidence below 60\%, 
and that number increases to 27\% when $\sigma/\mu \geq 1$.
The median confidence drops from 96\% when $\sigma/\mu < 0.55$ to 63\% when $\sigma/\mu \geq 2$.
In all cases, the median error remains below $33\%$ (and below 5.5\% for $\sigma/\mu \leq 0.5$). 
For all cases, the calculated $R_{IO}$ is wrong by less than 10\%. 

\begin{figure}[htp] 
    \centering
    \includegraphics[width=0.49\columnwidth]{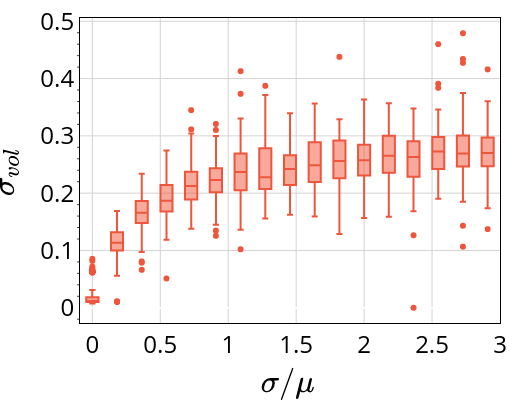}
    \includegraphics[width=0.48\columnwidth]{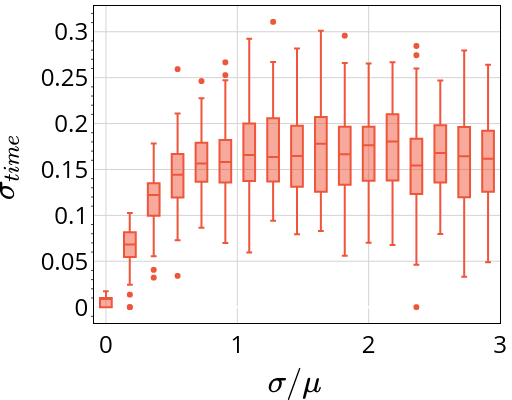}
    \caption{\stdvol and \stdtime from the experiments in \cref{fig:limit_cpusigma}.
    The term $\sigma/\mu$ on the x-axes represents the standard deviation of $t_{cpu}$ divided by the mean of $t_{cpu}$.}
    \label{fig:std_metrics}
\end{figure} 
\cref{fig:std_metrics} presents the metrics \stdvol (left) and \stdtime (right) for this experiment. 
As shown, both increase as the I/O variability increases (i.e., less periodic signal). 
Their variability for each point from the x-axis matches the variability observed in the error 
shown in \cref{fig:limit_cpusigma}. The median
periodicity score (see Sec.~\ref{subsec:additional_characterization}) 
is $98\%$ for $\sigma=0$, then drops to $67\%$ for $\sigma/\mu=0.55$, and to $57\%$ for $\sigma/\mu=2$.
Hence, when designing a technique (e.g., I/O scheduling) that uses the period obtained with \approach, 
one can study the robustness of their technique according to the values of \stdvol and \stdtime 
to decide on thresholds for these metrics, since some approaches will tolerate higher detection errors than others.

%% file: 03_00_01_limits_methodology.tex
\paragraph*{Methodology}

We have created ``semi-synthetic'' traces to allow for an extensive evaluation: 
First, we traced IOR\cite{ior} runs that represent a single I/O phase. Then, 
we generated application traces by combining I/O phases with a given amount of ``idle'' (no I/O) time between them. 
IOR was executed $100$ times on the \plafrim cluster
using 32 processes on four nodes. Each of them writes a $3.5$~GB file in $1$~MB contiguous requests.
One I/O phase was filtered out for being too long compared to the others (due to contention in the system), leaving 99 traces with an average duration of $10.4$~s ($\approx10$~GB/s), all inside $[10.22, 13.34]$~s.

An application is considered to be a sequence of $J$ non-overlapping iterations. 
Each iteration $j\leq$ $J$ has a compute phase of length $t_{cpu}^{(j)}$ 
followed by an I/O phase (of length $t_{io}^{(j)}$) where each of the $P$ processes writes an amount of data 
$v$ to the file system. 
The trace is created by selecting $J$ 
and $P$, and then, for each $j\leq$$J$, by:
\begin{itemize}
    \item Drawing $t_{cpu}^{(j)}$ from a normal distribution $\mathcal{N}(\mu,\sigma)$ truncated to only select positive values (with $\mu$ and $\sigma$ denoting the mean and standard deviation of $t_{cpu}$, respectively);
    \item Randomly picking one of the I/O phase traces, which consists of $P$ per-process traces;
    \item For each process  $k\in [1,P]$, adding a time $\delta_k$ at the beginning of its trace (without I/O). 
    $\delta_k$ is drawn from an exponential distribution of average \meanshift.  
    Process $0$ has $\delta_0=0$ to keep the boundaries of the I/O phase.
\end{itemize}

As the I/O phases' length depends on $\delta_k$, 
it allows us to represent both desynchronization between processes and I/O variability. 
Finally, for experiments with noise, we generated $200$ traces from IOR on a single process in two configurations:
low noise of nearly $500$~MB/s and high noise of nearly $1$~GB/s. The noise traces have 10 
periods of approximately $2.2$~s each. Noise is emulated by randomly selecting a sequence of noise traces 
and adding them to the application trace. 

\begin{figure}[tbp]
  \centering
  \begin{subfigure}[t]{0.48\textwidth}
    \includegraphics[width=\textwidth]{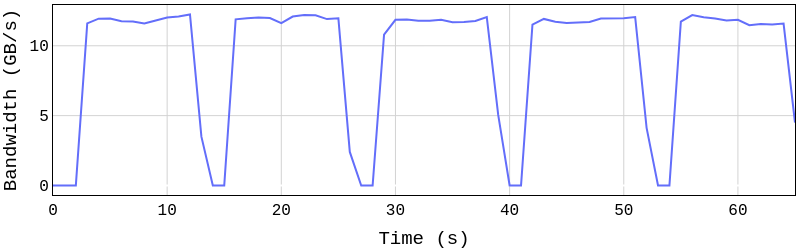}
      \caption{$t_{cpu}$ is 1/4 the duration of the I/O phase ($t_{io}$). }
    \label{fig:limit_dummy_1}
  \end{subfigure}\hfill
  \begin{subfigure}[t]{0.48\textwidth}
    \includegraphics[width=\textwidth]{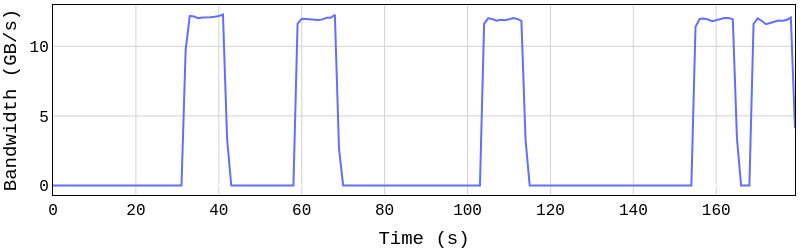}
      \caption{\new{$t_{cpu} \sim \mathcal{N}(11,22)$.}}
    \label{fig:limit_dummy_2}
  \end{subfigure}\hfill
  \begin{subfigure}[t]{0.48\textwidth}
    \includegraphics[width=\textwidth]{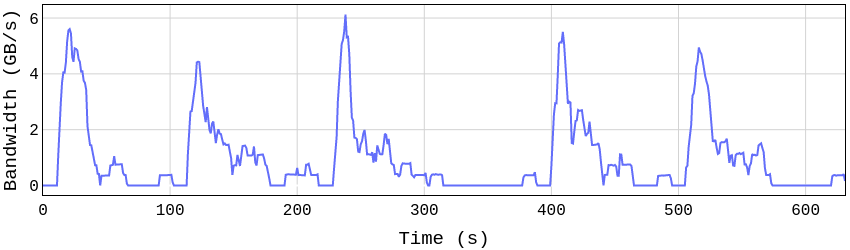}
    \caption{\new{$\bar{\delta}_k~\text{in I/O phases is 22 s}$.}}
    \label{fig:limit_dummy_3}
  \end{subfigure}
  \caption{Examples of ``semi-synthetic'' application traces.}
  \label{fig:app_example}
\end{figure}

For all experiments in this part, 
we used $f_s=1$~Hz, $P=32$ (the number of processes used for IOR), and $J=20$ 
(to be able to induce enough variability in each trace).
Figure \ref{fig:app_example} illustrates traces created with this approach.
We generate $100$ traces per parameter combination. 
For each one, we compute $\bar{T}$ the average period length 
and $T_d$ the period obtained with \approach. 
Finally, we calculate the \emph{detection error} as $|T_d - \bar{T}|/\bar{T}$. 
Note that $\bar{T}$ can only be computed using information from the trace generation, 
as the boundaries of I/O phases are not typically available. 
Reaching a low error means \approach provides a value that is close to the average period.

%% file: 03_00_02_error_figures.tex
\begin{figure*}[bt]
  \centering
  \begin{subfigure}[t]{0.25\textwidth}
    \includegraphics[width=\textwidth]{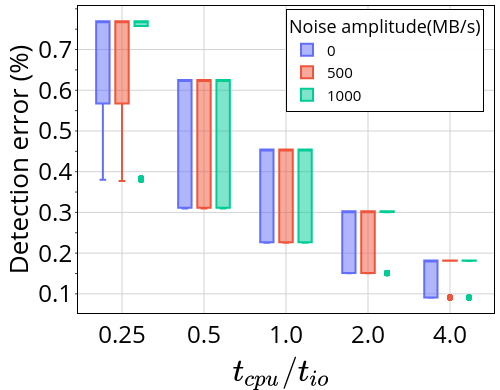}
      \caption{\new{... time between I/O phases (relative to their length) and noise.}}
    \label{fig:limit_meancpu}
  \end{subfigure}\hfill
  \begin{subfigure}[t]{0.36\textwidth}
    \includegraphics[width=\textwidth]{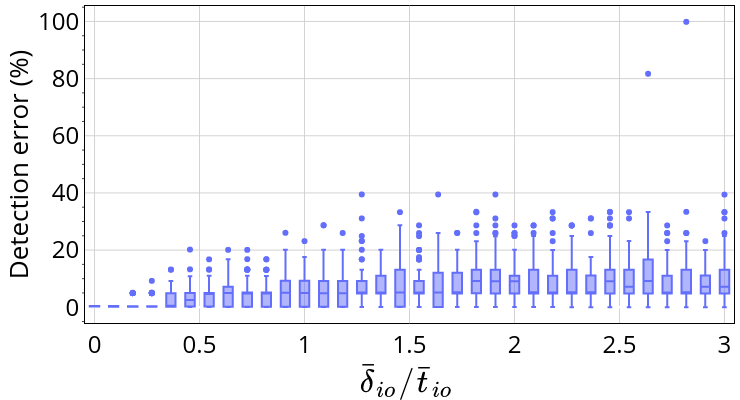}
      \caption{\new{...  \meanshift added to processes' I/O phases.}}
    \label{fig:limit_shift}
  \end{subfigure}\hfill
  \begin{subfigure}[t]{0.36\textwidth}
    \includegraphics[width=\textwidth]{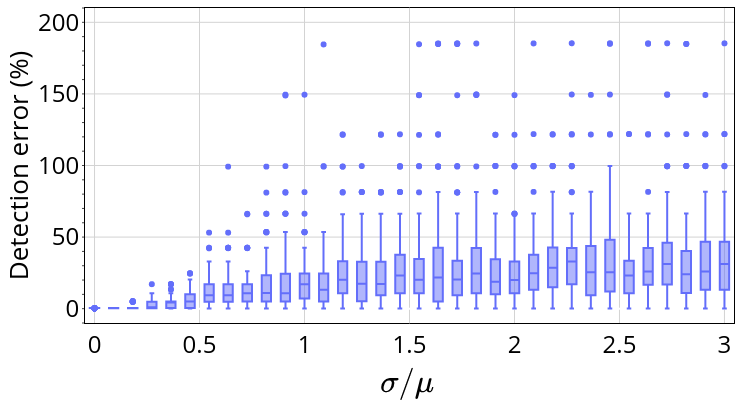}
    \caption{\new{... variability of time between I/O phases.}}
    \label{fig:limit_cpusigma}
  \end{subfigure}

  \caption{Detection error as a function of the ...}
  \label{fig:detection_error}
\end{figure*}

%% file: 03_02_results_validation.tex
\subsection{Case Studies}
\label{sec:results_validation}
\label{sec:results}
In \cref{subsec:autocorrelation}, 
we showed the scalability of \approach using  IOR\cite{ior}. 
To evaluate our method further, in this section we:
\begin{inparaenum}[(a)]
\item analyze a real application (LAMMPS\cite{LAMMPS}) with low I/O bandwidth, 
\item demonstrate the compatibility of \approach with a Darshan profile of Nek5000\cite{nek5000}, and 
\item use a mini-app (HACC-IO\cite{haccio}) with high I/O bandwidth to
highlight the detection and prediction capabilities of \approach.
\end{inparaenum}
%
The experiments (a) and (c) were performed on the \machine cluster, where
a typical node has 96 cores, and the access mode is user-exclusive. The shared
file system (IBM Spectrum Scale) has a peak performance of 106~GB/s for writes and
120~GB/s for reads.

\begin{figure}[tbp]
    \centering
	\includegraphics[width=\columnwidth]{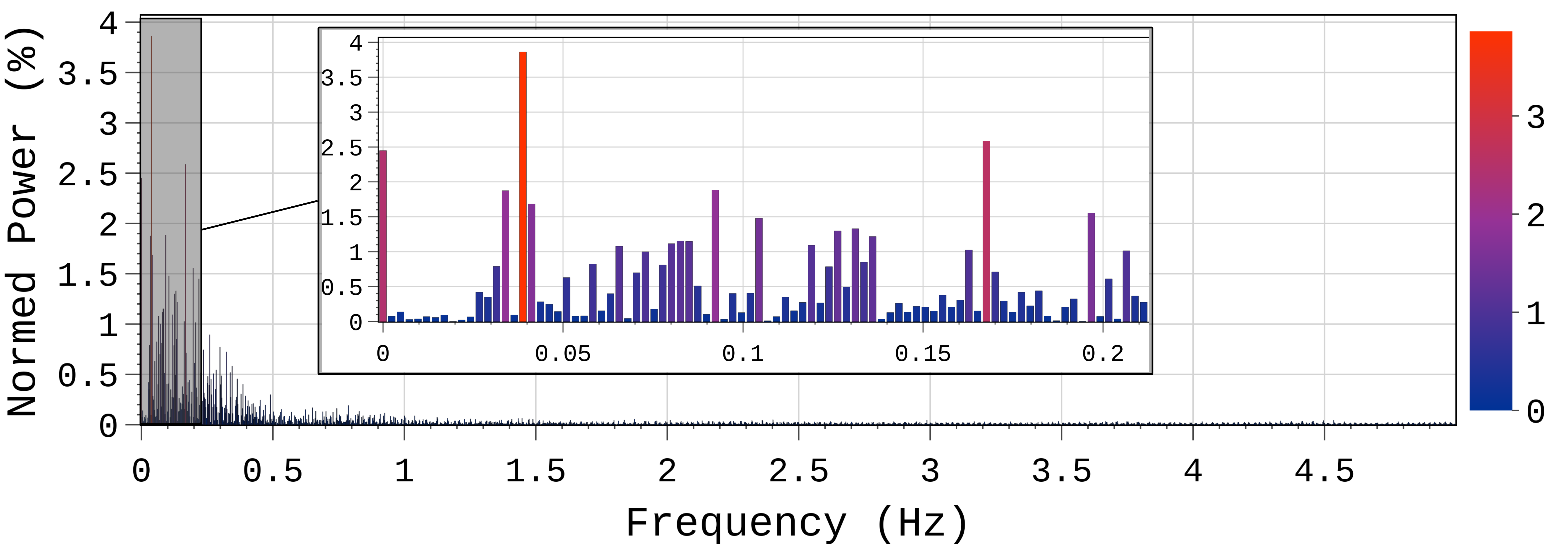}\\[0.2cm]
\includegraphics[width=.95\columnwidth]{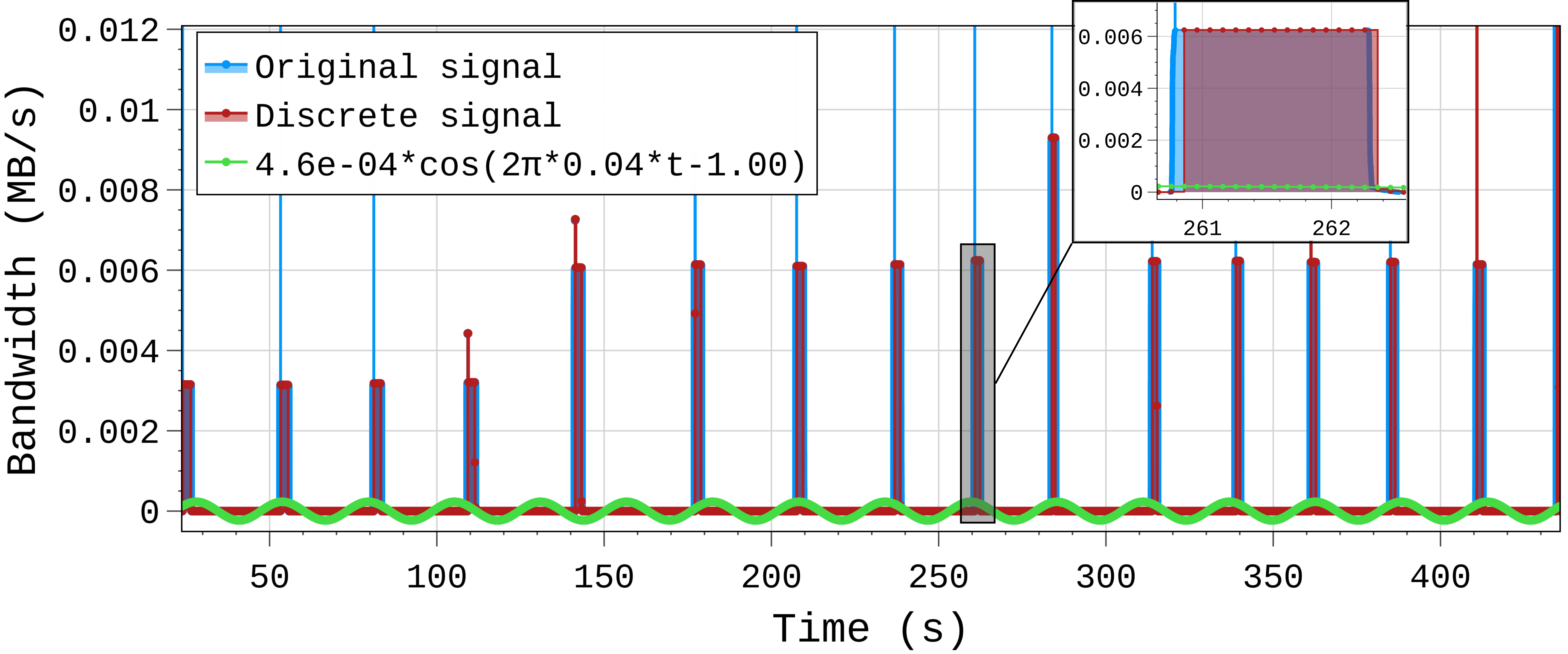}
    \caption{Result of \approach on LAMMPS with 3072 ranks: normed power spectrum (top) and time behavior (bottom).}
    \label{lammps}
\end{figure}

%
\paragraph{\textbf{Real application with low I/O bandwidth}}
We demonstrate our approach on LAMMPS~\cite{LAMMPS} with 3072 ranks. 
We use the 2-d LJ flow simulation with 300 runs that dumps all atoms every 20 runs. 
Using \approach with $f_s=10$~Hz in the detection (offline) mode, the result was obtained in 2.2~s. The 
top part of \cref{lammps} shows the single-sided power spectrum.
As illustrated, \approach found a single dominant frequency at 0.039~Hz (25.73~s) 
with a confidence of 55.0\%. Due to the moderate contribution 
of the frequency at 0.16~Hz (5.9~s), the low confidence is justified. 
Using autocorrelation (for an additional cost of 0.26~s), the confidence 
is refined and we obtain 84.9\%  (with only a single peak detected at 25.6~s).  
For comparison, the real mean period for this execution was 27.38~s. 
The bottom part of \cref{lammps} demonstrates the result of \approach in the time domain, 
which shows the low I/O performance due to the writing method. 
As observed, the dominant frequency does not perfectly fit all phases (e.g., at 143~s), 
justifying again the low confidence obtained. 
Still, it provides an adequate and concise representation of the temporal I/O behavior of the application, 
which is what we aimed for. 
Note that the results can be improved by adapting $\Delta t$, 
as the next example shows. 
Still, the offline approach demonstrated here could be used, for example, to 
feed an I/O scheduler at the start with the period of the I/O phases from previous executions. 
%
%
%

\begin{figure}[bhtp]
    \centering
	\includegraphics[width=\columnwidth]{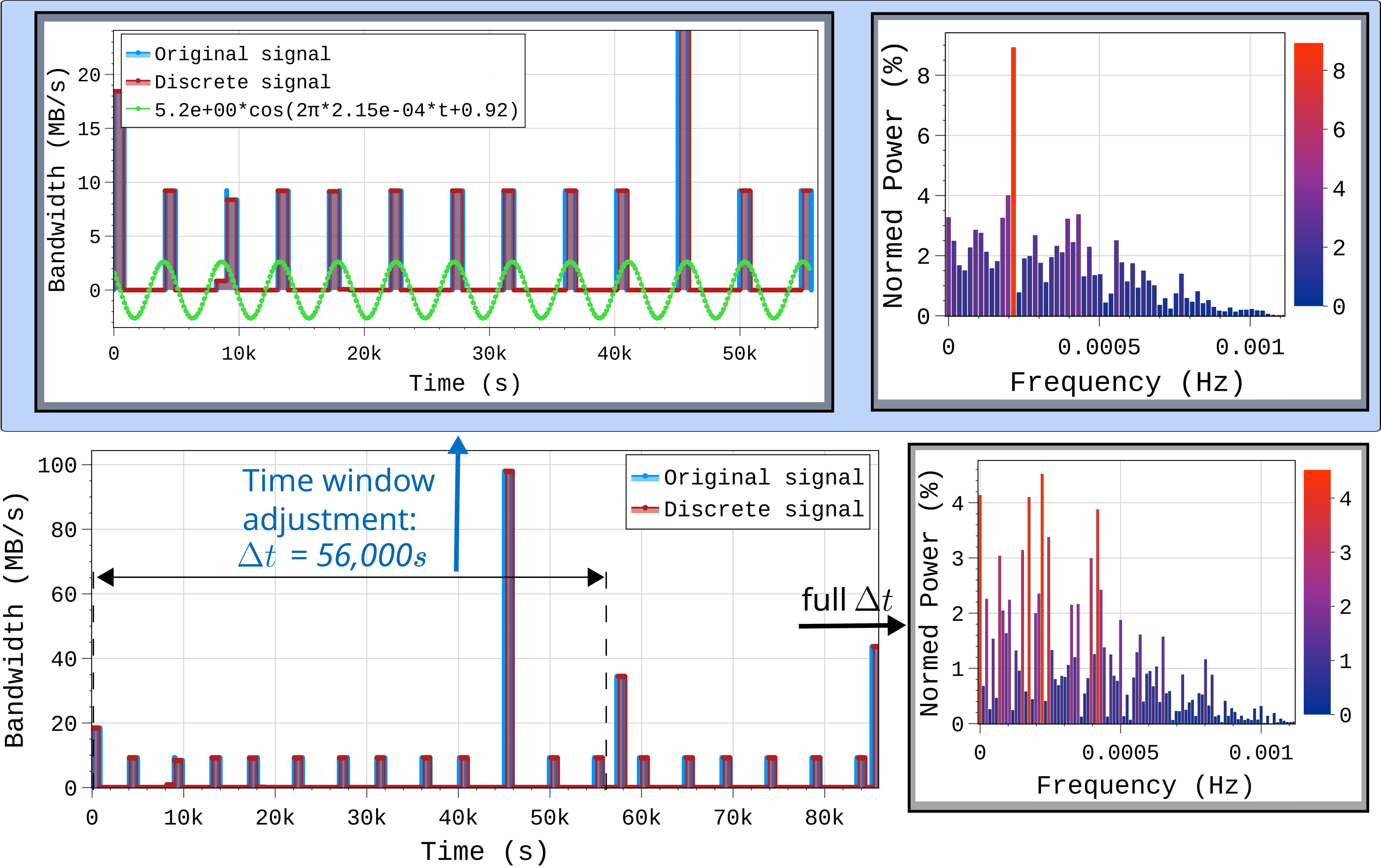}
    \caption{\new{Result of \approach on Nek5000 with the full trace (bottom), 
	and a reduced time window $\Delta t= 56,000$~s (top).
 }}
    \label{nek5000}
\end{figure}

\paragraph{\textbf{Compatibility with other tools}}
%
%
%
For this example, we downloaded from the \emph{I/O Trace Initiative} website\cite{MotiBVD0CAS23} a Darshan profile\footnote{\url{https://hpcioanalysis.zdv.uni-mainz.de/trace/64ed13e0f9a07cf8244e45cc}}
of Nek5000~\cite{nek5000} (turbulence simulation) executed with 2048 ranks on the Mogon II cluster.
\approach extracted the heatmap from Darshan profile and automatically set the sampling frequency to the bin widths in seconds ($f_s=0.006$~Hz).  While a higher sampling frequency could be used here, there 
is no advantage due to the constant behavior in the bins. 
\approach detected that the I/O phases are not periodic if the entire trace is considered ($\Delta t =86,000$~s) 
as shown in the lower part of \cref{nek5000}. 
%
%
This is due to the irregular I/O phases at roughly 57,000 and 85,000~s, 
which write each around 30~GB, compared to the 7~GB of the remaining ones (except the phases at 0~s and 45,000~s, which write 13 and 75 GB, respectively).
Moreover, the bins that write 7~GB are not equally spaced. 
However, if the time window is set to $\Delta t =56,000$~s, 
\approach detects a period of 4642.1~s with a confidence of 85.4\% as the upper left part of \cref{nek5000} shows. Moreover, 
the power spectrum is less noisy compared to the previous case shown on the right side of \cref{nek5000}, 
where we zoomed to the relevant frequencies. 
Consequently, a clear outlier can be detected.
In the next example, we demonstrate how the online prediction automatically 
adapts $\Delta t$ to improve the prediction results. 

\paragraph{\textbf{Detection and prediction with high I/O bandwidth}}
%
%
Next, we use HACC-IO~\cite{haccio},
which mimics one I/O phase of HACC (Hybrid/Hardware Accelerated Cosmology
Code)~\cite{SH12}. 
HACC-IO has four steps: compute, write, read, and verify. 
We added a loop around these steps to execute them periodically. 
Moreover, at the end of each loop iteration, we added a single line to flush the collected data out to the trace file. 
On a login node, we deployed FTIO in the online prediction mode.
We executed this example with 3072 ranks on the \machine cluster. 

\begin{figure}[bhp]
    \centering
\includegraphics[width=\columnwidth]{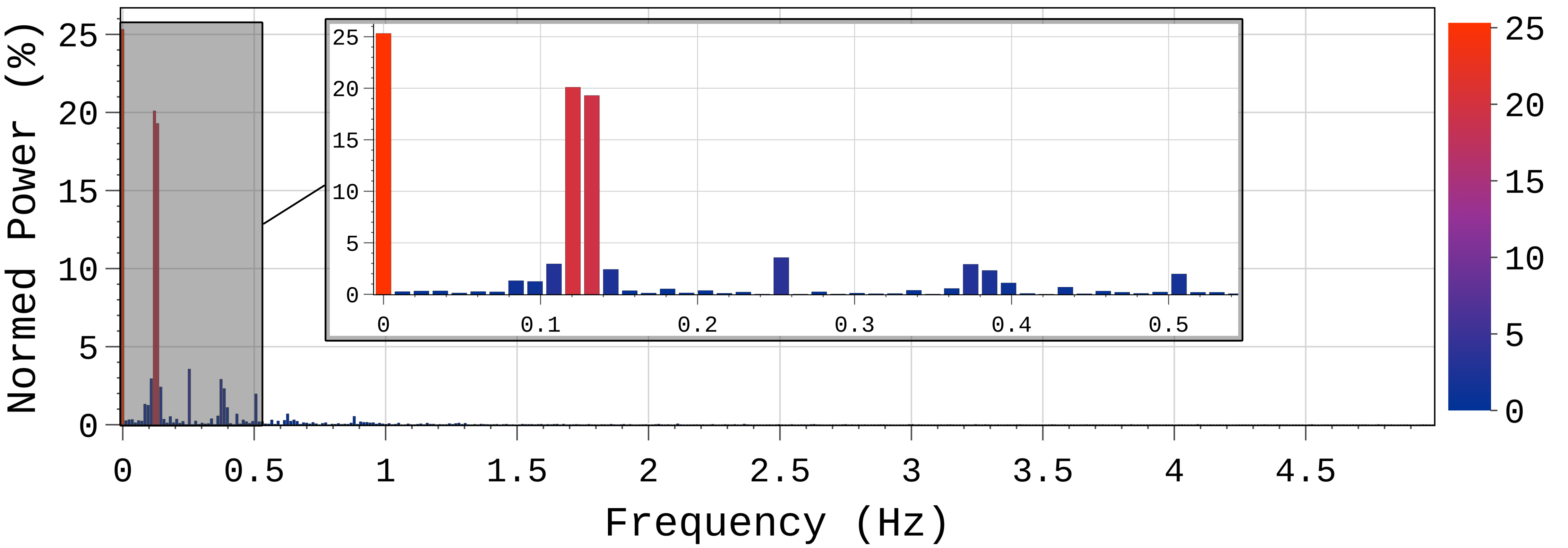}
    \caption{\new{Single-sided power spectrum from DFT on HACC-IO with 3072 ranks and a sampling frequency $f_s=10$~Hz.}}
    \label{fig:hacc_3072_detect_half_spectrum}
\end{figure}

\begin{figure}[tp]
    \centering
    \includegraphics[width=\columnwidth]{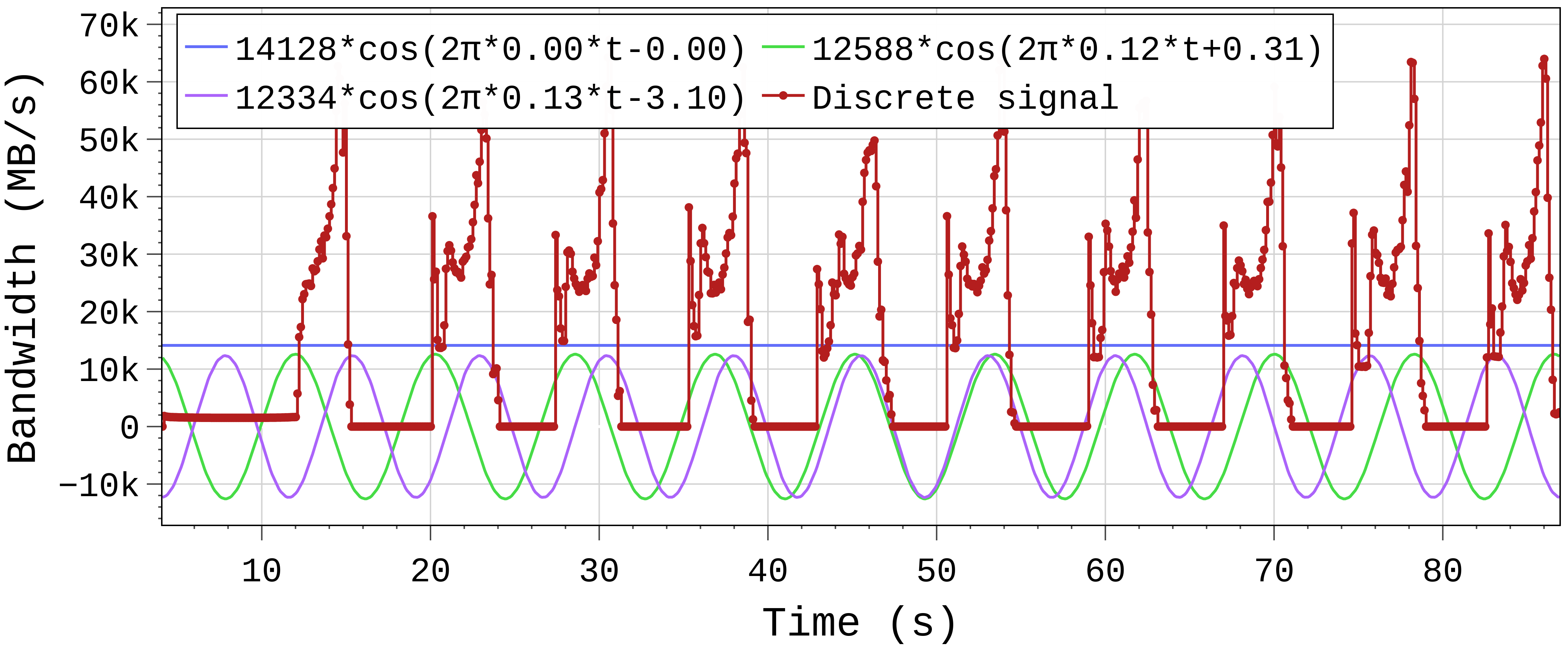}
    \caption{DC Offset and the highest-contributing frequencies found in the temporal behavior of HACC-IO with 3072 ranks executed on the \machine cluster.}
    \label{fig:hacc_3072_detect}
\end{figure}

\textit{1) Offline evaluation:}
We first look into the output of the offline evaluation performed over the whole
trace after the end of the execution. \cref{fig:hacc_3072_detect_half_spectrum} presents 
the single-sided normed power spectrum.
Two candidates for the dominant frequency were found:
0.1206~Hz ($c_k =51$\%) and 0.1326~Hz ($c_k =48.9$\%). 
As the former one has the highest contribution, it is the dominant frequency corresponding to a period of $8.29$~s. 
Note that the application is by design periodic. 
However, if we study qualitatively the execution in~\cref{fig:hacc_3072_detect}, 
we see that 
the first I/O phase was significantly delayed: it lasts from $4.1$~s to $15.3$~s. 
This changing behavior results in a less periodic signal and explains our moderate
confidence. 
Indeed, the average period 
is $8.7$~s, which becomes $7.7$~s  
without the first I/O phase. \cref{fig:hacc_3072_detect} shows the top three signals found by DFT. 
As presented, the I/O phases align more with the 0.120~Hz signal (green) at the start and 
with the 0.132~Hz signal (purple) near the end. 

\begin{figure}[h]
    \centering
    \includegraphics[width=\columnwidth]{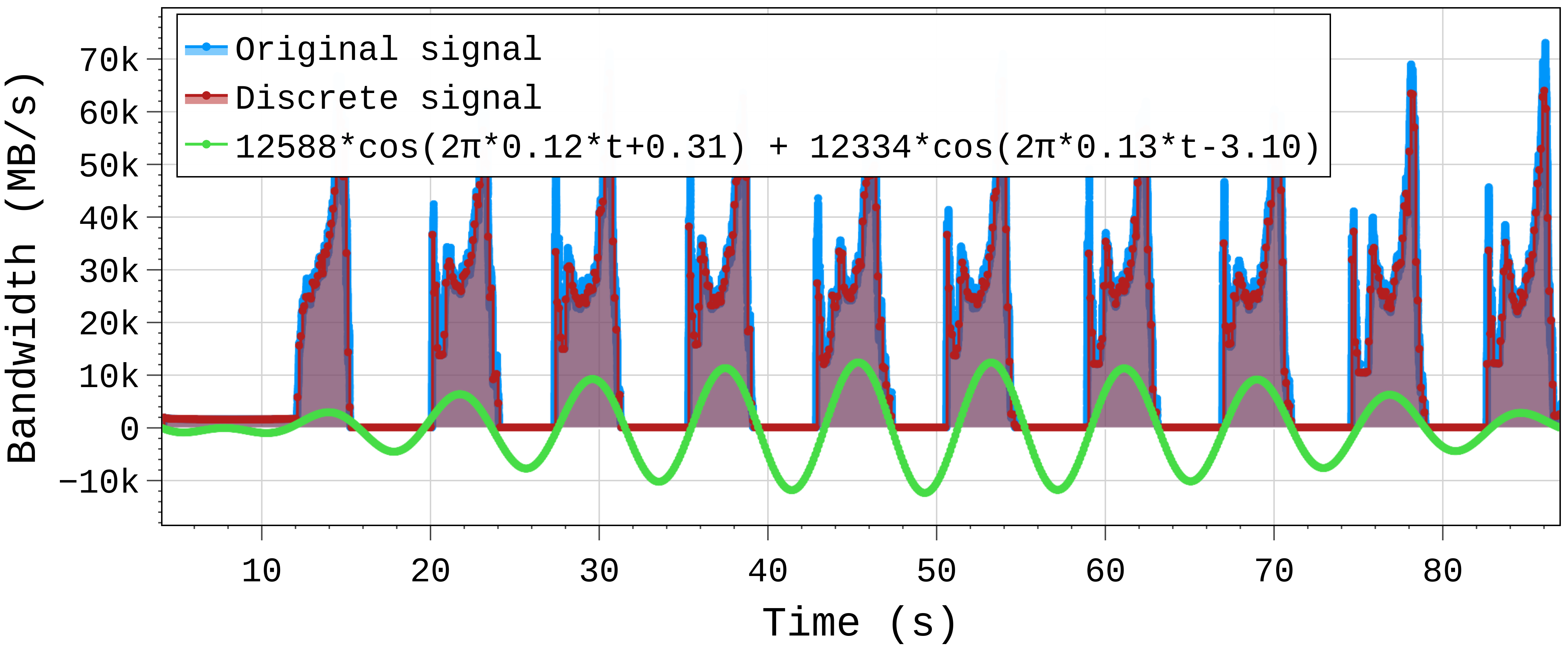}
    \caption{HACC-IO with 3072 ranks. By combining the dominant frequency candidates, a more accurate description of the application's temporal I/O behavior can be obtained. However, this is more difficult to interpret, and hence, not used here.}
    \label{fig:hacc_3072_merged_freq}
\end{figure}

As the two candidates for the dominant frequency have very close contributions and are consecutive, 
one approach could be to merge them by taking the sum of their cosine waves. 
This is shown in \cref{fig:hacc_3072_merged_freq}, and would provide a more accurate representation of the application's temporal I/O behavior. 
However, in this paper, we focus on representing the behavior with a single period, which is concise and 
can easily be used as an input for techniques such as I/O scheduling. 
In contrast, a more detailed application profile could include several dominant frequency candidates and their 
contributions.
We plan on exploring such profiles in the future. 
\new{Note that as the first phase is often prolonged due to initialization overheads, \approach provides an option to skip it. 
Next, we show how the online version of \approach automatically handles changing I/O behavior.
}

\begin{figure}[tbhp]
\centering
\newcounter{i}
\newcounter{j}
\subfloat[
    \new{Above the time axis is the ground truth: 
    the blue rectangles are the I/O phases, and 
    above them is the time between their start times. Below the time axis are the
predictions: at time 63~s, we predicted a period of 8~s based on the history up to time 23~s (thick red line).}\label{fig:hacc-online}]{
\resizebox{.98\linewidth}{!}{
  \begin{tikzpicture}
    \begin{scope}[xscale=1/8]
    \draw[->,thick] (-0.5,0) -- (90,0) node[below] {Time};
      \foreach \x in {0,10,20,...,80}{
		  \draw[-,black] (\x,-.1) -- (\x,.1);			
		  \node at (\x,-0.1) [below] {$\x$};
      }

      \foreach \x in {4.1  , 20.0 , 27.3 , 35.2 , 42.8 , 50.5 , 58.8 , 66.9 , 74.5 , 82.5}{
		  \stepcounter{i}
		  \node (s_\thei) at (\x,0) {};
      }
     \setcounter{i}{0}
     \foreach \x in {15.3 , 24.1 , 31.3 , 39.2 , 47.4 , 54.7 , 63.1 , 71.1 , 79.0 , 86.9}{
		  \stepcounter{i}
		  \node (e_\thei) at (\x,0) {};
      }

      \foreach \x in {1,...,10}{
		  \draw[fill=blue!20] ($(s_\x)$) rectangle ($(e_\x.north)+(0,0.1)$);
      }

	\setcounter{j}{1}
	\setcounter{i}{2}
      \foreach \x in {15.9 , 7.3  , 7.9  , 7.6  , 7.7  , 8.3  , 8.1  , 7.6  , 8.0}{
		  \draw[-,black] ($(s_\thej.north)+(0,.1)$) -- ($(s_\thej.north)+(0,.3)$);
		  \draw[-,black!80] ($(s_\thej.north)+(0,.2)$) -- ($(s_\thei.north)+(0,.2)$) node[midway,above] {\x} ;
		  \stepcounter{i}\stepcounter{j}
      }
	  \draw[-,black!40] ($(s_\thej.north)+(0,.1)$) -- ($(s_\thej.north)+(0,.3)$);

	\setcounter{i}{0}
 	  \node at (5,-1.8) {\texttt{pred}:};
      \foreach \x/\y/\p in {4.1/15.3/11.1 , 4.1/24.1/9.9 , 4.1/31.3/9 , 4.1/39.2/8.7 , 23/47.4/8.1 , 23/54.7/7.9 , 23/63.1/8 , 47/71.1/8 , 47/79.0/7.9 , 47/86.9/8}{
		  \draw[->,black!40] ($(\y,-0.5-0.1*\thei)$) -- ($(\x,-.5-0.1*\thei)$) ;
		  \draw[-,black!40,dashed] ($(\y,0)$) -- ($(\y,-1.6)$) node[below,black] {$\p$};
		  \stepcounter{i}
      }
      
          \draw[->,black,very thick] ($(47.4,-0.5-0.1*4)$) -- ($(23,-.5-0.1*4)$) ;
          \draw[-,black,dashed,very thick] ($(47.4,0)$) -- ($(47.4,-1.6)$) node[below,black] {$8.1$};

		  \draw[->,red,very thick] ($(63.1,-0.5-0.1*6)$) -- ($(23,-.5-0.1*6)$) ;
		  \draw[-,red,dashed,very thick] ($(63.1,0)$) -- ($(63.1,-1.6)$) node[below,red] {$8$};
	
    \end{scope}
   \end{tikzpicture}
}
}\\[0.2cm]
\subfloat[Details of the $7^{th}$ prediction (thick red line from above).]{\includegraphics[width=.9\columnwidth]{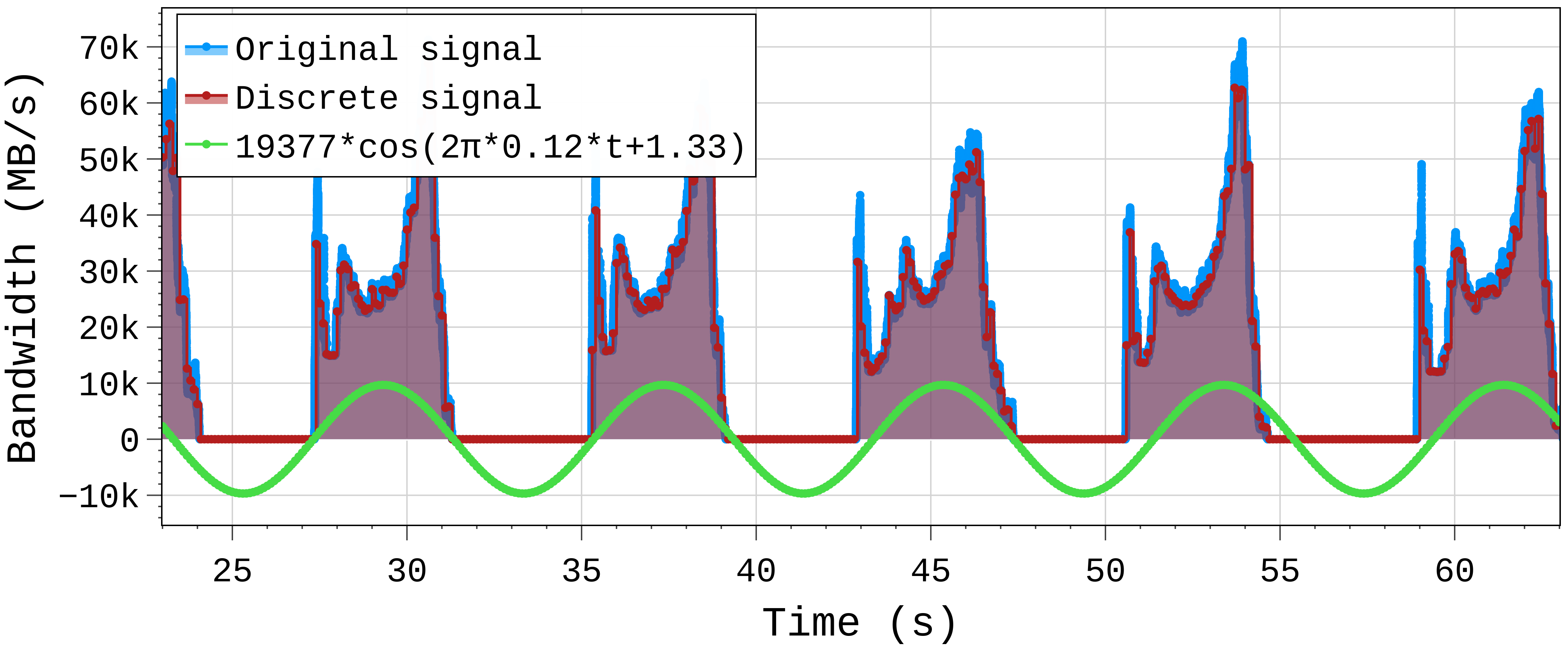}}
\caption{Online prediction during the execution of HACC-IO with 3072 ranks.}
\end{figure}

\textit{2) Online Prediction:}
As discussed in Section~\ref{sec:param}, the time window for \approach predictions can be 
automatically adapted according to the found frequency. 
For the ten I/O phases, which started on average every 8.7~s, the average
obtained period is 8.66~s. All predictions are shown visually in
\cref{fig:hacc-online}.
%
%
As shown,  
predictions were done at the end of each I/O
phase (dashed vertical lines) when new data became available. 
At the end of the $3^{rd}$ prediction, a
dominant frequency was identified for the third time. As the 
$4^{th}$ prediction had already started, that information was made
available for the $5^{th}$ prediction.
Hence, for the next evaluation, we only kept the data between 
23.1~s ($47.4-3\times 8.1=23.1$~s) and 47.4~s (time of
the $5^{th}$ prediction). In the figure, this is represented in bold. Similarly,  at 
$8^{\text{th}}$ prediction, the time window was again adapted ($71.1-3\times 8 = 47.3$~s). 
In this section we 
demonstrated the use of 
\approach with large-scale applications for both offline detection
and online prediction 
alongside metrics that gauge the confidence
in our results.
I/O variability and changing behaviors 
often caused the signal to be less periodic, resulting 
in a moderate confidence in the obtained results. The observation 
from HACC-IO indicates that, in these situations, the online 
prediction approach can yield the best results by adapting the time window.

%% file: 03_01_overhead.tex
\subsection{\ready{Overhead of the Tracing Library}}
\label{subsec:overhead}
The tracing library can be used for offline detection or online prediction. From those,
we examine the online approach as it has a higher overhead since it sends information to the file more often (see \cref{subsec:gathering_io}). To measure it, we executed 
IOR with the same settings as in \cref{subsec:anomaly}, on the \machine cluster (see \cref{sec:results}) with different numbers of processes (all multiples of 96 as we have 96 cores per typical node).

\begin{figure}[tp]
    \centering
    \includegraphics[width=\columnwidth,trim= 0 0 0cm 0,clip]{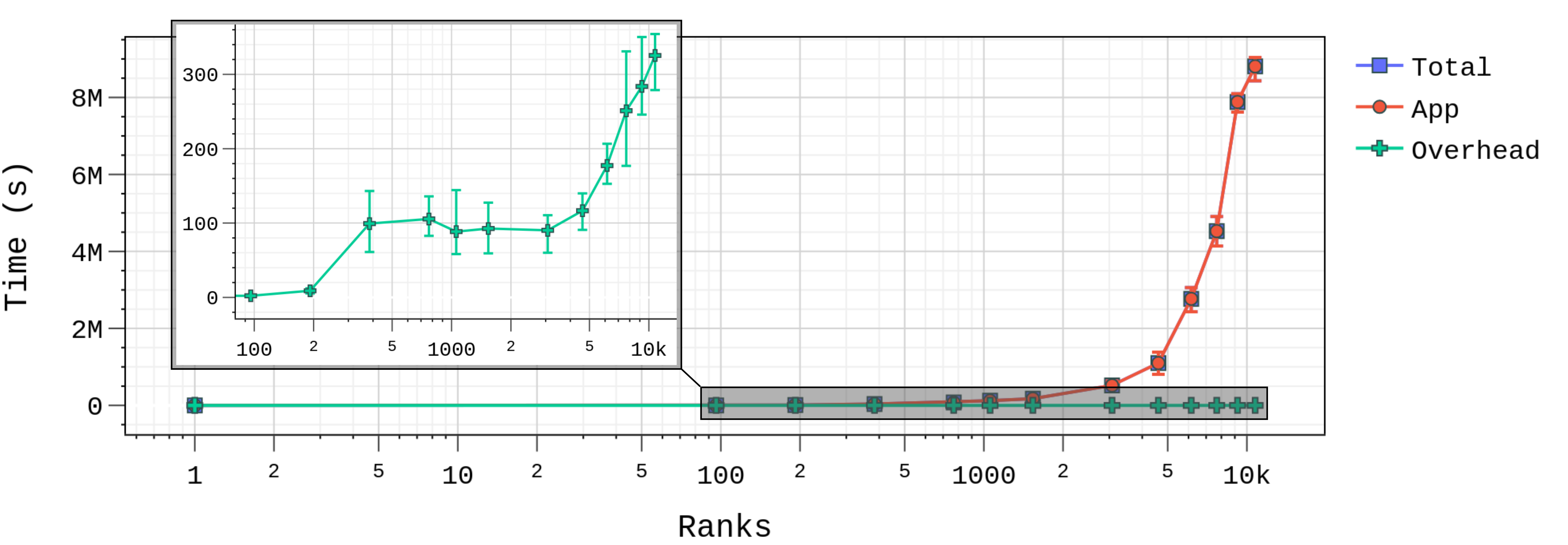}\\[0.1cm]
    \includegraphics[width=\columnwidth,trim= 0 0 0cm 0,clip]{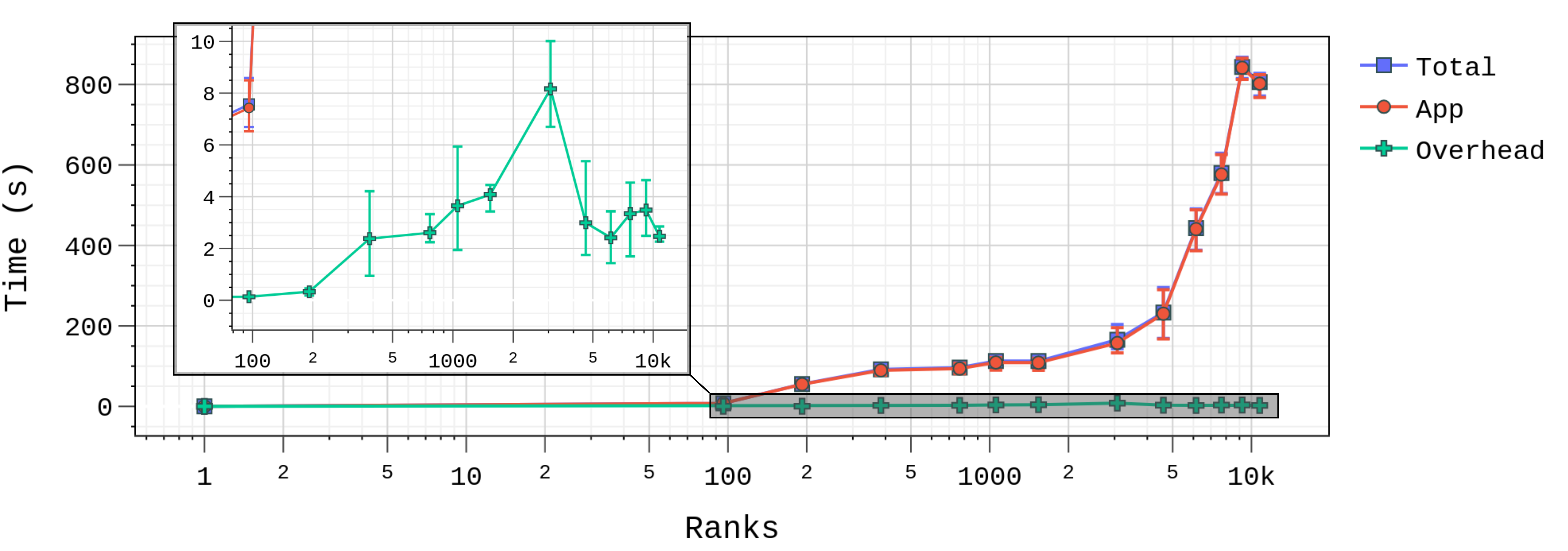}
    \caption{Overhead of our approach across different rank configurations (from 1 till 10752 ranks). The top shows the aggregated time across all ranks, while the bottom shows the time of MPI rank 0 only.}
    \label{fig:overhead}
\end{figure}
\cref{fig:overhead} shows the overhead of the tracing library, with the top part showing the aggregated time, while the bottom plot shows the time from the MPI rank 0 perspective. The numbers of ranks on the x-axis are in log scale, and the sum of the application time (App) and overhead is the total time. 
More precisely, we instrumented our library calls and subtracted the corresponding values from the measured total time to derive the application time.
As observed,
for capturing and logging the I/O data, our tracing library has a low overhead:  
a maximum of 0.6\% for the aggregated overhead 
and 6.9\% for the overhead for rank 0 only. 
The data gathering from the different ranks is the major source of overhead. 
For comparison, in the same configurations,
the overhead of the offline approach 
ranged from 0.78~s (0.13\%) at 96 ranks to 50.9~s (0.004\%) at 4608 ranks in the aggregated overhead time and increased nearly linearly from   
0.065~s (1.03\%) to 3.84~s (1.58\%) for the overhead for rank 0 only. 

It is important to notice that our approach can be used with other data collection strategies (see \cref{subsec:gathering_io}) 
which would have different implications 
as demonstrated in the second example in \cref{sec:results}. 
The execution time of the analysis is of minor importance, as mentioned, and depends on the length of the time window.
For all examples in this paper, the longest analyses took: 
2.2~s for LAMMPS,  5.7~s (5.9~s with autocorrelation) for IOR, 
8.7~s (8.5~s with adjusted $\Delta t$) for Nek5000 of which pyDarshan consumed 3.8~s to import the data, 
and 3.6~s for the offline detection for HACC-IO.

%% file: 04_application.tex
\section{Use Case: I/O Scheduling}
\label{sec:motivation}

Although work on I/O scheduling~\cite{boito2023io,benoit:hal-04038011,jeannot2021scheduling,dorier2014calciom} has proved how critical the knowledge of I/O phases is, \approach is the first runtime solution that provides simple and lightweight access to this information. To illustrate the utility of \approach, we coupled it with Set-10~\cite{boito2023io}, an I/O scheduling heuristic. The goal of Set-10 is to mitigate file-system contention, exploiting that the frequencies at which jobs perform their I/O usually differ. For this purpose, Set-10 groups jobs according to their I/O period. 
It then grants shared file-system access to different groups (based on
priorities) and mutually exclusive access to individual jobs within the
same group. In case FTIO is used together with Set-10 (later denoted as "Set-10 + FTIO"), the
priorities for the groups (i.e., the sets) are calculated based on the period $T_d$ provided by FTIO.
According to the Set-10 algorithm, applications with the smallest period receive the highest priority and, therefore, most of the bandwidth \cite{boito2023io}.
Note that in the original Set-10 implementation, priorities for the groups were calculated using
the characteristic time $w_{iter}$ (i.e., the average time between the beginning of two consecutive I/O phases) as described in~\cite[Section IV-C]{boito2023io}. 

\begin{figure}[tbp] 
    \centering
    \includegraphics[width=0.325\columnwidth]{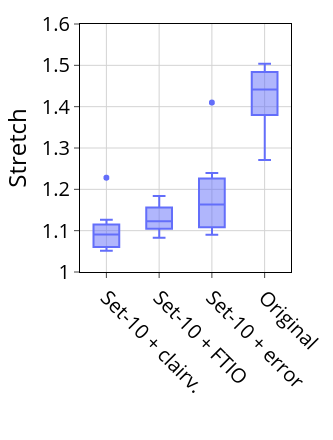}
    \includegraphics[width=0.325\columnwidth]{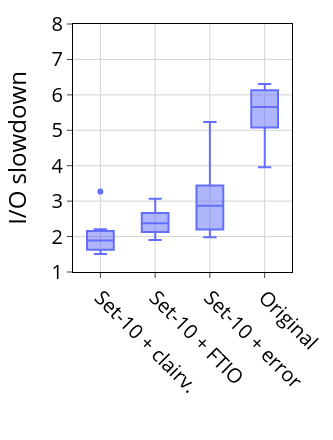}
    \includegraphics[width=0.325\columnwidth]{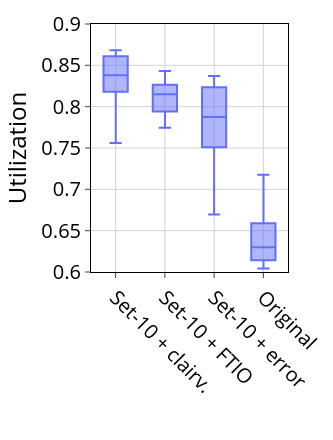}
    \caption{
    \new{Comparison of clairvoyant Set-10, Set-10 with \approach, Set-10 with 50\% error injected to the FTIO-provided periods, and the original configuration without Set-10. 
    The figures show the stretch 
    (how much slower jobs were compared to running in isolation: lower is better), 
    the I/O slowdown 
    (how much slower I/O was compared to isolation: lower is better), 
    and the utilization 
    (how much of the time was NOT spent on I/O: higher is better). 
    The boxplots (with 1.5*IQR whiskers) group ten executions. The y-axes do not start at zero and are all different.} 
    }
    \label{fig:iosets_results}
\end{figure}

For our experiments, we used an implementation, described in~\cite{iosets_techreport}, of Set-10 on BeeGFS and deployed \approach to determine each job's period at runtime. Our workload consists of one high-frequency and 15 low-frequency applications derived from the IOR benchmark. They were designed to include, in isolation, periods of 19.2~s (high frequency) or 384~s (low frequency), with I/O consuming 6.25\% of each period.
\cref{fig:iosets_results} shows the results, comparing four situations:
\begin{itemize}
    \item ``Set-10 + clairv.'' is a clairvoyant application of the scheduling heuristic, meaning that the \emph{ideal} (in isolation) periods (19.2 or 384 s) are provided manually in advance. 
    \item  ``Set-10 + \approach'' combines the heuristic with \approach, which determines the \emph{actual} periods at runtime. In this case, Set-10 uses the most recent prediction from \approach. 
    \item  ``Set-10 + error'' uses predictions {\em worse} than \approach: the predictions given by \approach are randomly increased or decreased by a factor of 50\% before provided to Set-10.
    \item ``Original'' corresponds to BeeGFS without any modifications and serves as the baseline.
\end{itemize}
%

\new{We evaluate these algorithms with three metrics: the stretch, the I/O slowdown, and the system utilization.}
The \textit{stretch} quantifies the overall slowdown factor for an application caused by inter-job file-system interference; the \textit{I/O slowdown} represents the factor by which its I/O time was increased. 
\new{Thus, the lowest value of both metrics is 1.}
\new{Both are calculated by taking the geometric mean of all applications from each execution.} 
The \textit{Utilization} \new{($\in [0,1]$)} is a system metric that specifies how much of the node time was spent on computation instead of I/O. 
More details about these metrics are given in~\cite[Section V-D]{boito2023io}, and more on this experiment in~\cite{iosets_techreport}.

\new{
The results achieved with \approach are close to the clairvoyant version---only 2.2\% worse in stretch, 19\% in I/O slowdown, and 2.3\% in utilization. 
In contrast, the version where we inject errors to \approach results made stretch worse by 5\%, utilization by 4\%, and I/O slowdown 27\% higher, compared to the ``Set-10 + \approach'' version, in addition to presenting higher variability.
Thus, the main observation from \cref{fig:iosets_results} is that \approach hits a sweet spot in performance where a better prediction would not improve the performance observed by the system or the users, however, a worse prediction would increase the variability and impair the performance of the system. That indicates \approach provides results that are good enough for Set-10, and a more accurate method, if available, would not have much margin to improve the results significantly.
Compared to not using Set-10, the \approach-powered version decreased the mean stretch and I/O slowdown by 20\% and 56\%, respectively, and increased utilization by 26\%.} These results show how well \approach fills the knowledge gap, making the improvements that Set-10 allows possible in practice, where the period is not known in advance.

%% file: 05_related_work.tex
\section{\ready{Related Work} }
\label{sec:related}

Since I/O performance depends on many parameters~\cite{9229642,8752753,10.1145/3337821.3337902,10.5555/3323234.3323267,9355272},
profiling tools such as Darshan~\cite{carns200924,snyder2016modular} and other more holistic approaches~\cite{Lockwood2017UMAMIAR,xu2016lioprof} can be used by an expert to obtain insights about application I/O behavior and improve it. However, these large profiles are 
not easily automatically exploitable at run time by optimization techniques, which must focus on simpler metrics.
For example, in the context of cache management and I/O prefetching, it is useful to predict future I/O requests~\cite{10.1145/3458817.3476181}. That has been done using neural networks~\cite{10.1007/978-3-030-29400-7_4}, ARIMA time series analysis~\cite{1271185}, pattern matching~\cite{10.1007/978-3-319-09873-9_21}, context-free grammars~\cite{dorier:hal-01238103}, etc.
Although \approach could be used to predict future accesses, it is fundamentally different from these approaches because we focus on \emph{I/O phases}, not \emph{I/O requests}.
%
%
%
\new{Working at this higher level brings the challenge of not knowing when the I/O phases start and end (see \cref{sec:introduction}), particularly since the phases are logical groupings of I/O requests, not individual events. Still, the period of the I/O phases is a metric worth finding as it can be easily and directly exploited by contention avoidance algorithm, as demonstrated in \cref{sec:motivation}. }

\new{Aside from being able to handle changing I/O behavior (see \cref{subsec:online}), \approach can be executed online due to its low overhead (see \cref{subsec:overhead}). The main advantages of \approach in this comparison basically stream from the unique properties of DFT. 
Compared to popular machine learning (ML) approaches from the time domain, like neuronal networks (NN)~\cite{lucabez:hal-02276191} and LSTMs~\cite{8855713,LI19}, decision trees and other supervised methods~\cite{7776517}, 
or a combination of supervised and unsupervised techniques~\cite{9355272}, \approach, and in particular DFT, does not require a learning phase. 
Additionally, \approach does not require past system logs, different from recent regression-based approaches~\cite{Kim_Sim_Wu_Byna_Son_2023} and other strategies~\cite{8855713,LI19,WT18,9355272,7776517,Kim_Sim_Wu_Byna_Son_2023}, 
Moreover, compared to approaches that predict future I/O activity, such as ARIMA~\cite{1271185}, DFT does not require  defining several thresholds and parameter estimations.
In contrast, DFT yields the signal's frequency components rather than providing a detailed time model. Still, this is enough to predict the \emph{period} of the I/O phases, as demonstrated in \cref{sec:evaluation}, and in particular for the I/O scheduling use case (\cref{sec:motivation}).
%
%

Determining the application-level period from time models usually requires defining thresholds, which can be system and application-dependent (see \cref{sec:introduction}).  
Even if an approach, such as ARIMA, accurately predicted the time behavior shown in \cref{fig:exampleIO}, we would still need to analyze the result further to extract the period. In this context, finding suitable thresholds to detect the phases is challenging, primarily due to the varying nature of I/O (e.g., I/O burst, slow I/O, etc.). 
\approach directly utilizes DFT to overcome such challenges and, combined with outlier detection methods, determines the period of the I/O phases. 
Further characterization can be provided based on the identified period as described at the end of \cref{subsec:autocorrelation}. 
Additionally, as shown in \cref{sec:results}, the parameters of DFT allow \approach to adapt to changing behavior (using $\Delta t$) and to specify the range of interesting I/O (using $f_s$). Furthermore, the online time window adaptation usually decreases the overhead of \approach further, as less samples ($N$) are included in the analysis (see \cref{subsec:DFT}), making this approach even more favorable for online period prediction.
}

More general characterization efforts usually focus on aspects such as spatiality and request size~\cite{10.1145/2597652.2597686,osti1009541}, 
using information from MPI-IO~\cite{6546095,6217423,10.1145/1996130.1996138},  ML-based methods~\cite{lucabez:hal-02276191}, etc.
In contrast, FTIO focuses on the \emph{temporal} behavior (specifically on the periodicity), and hence is \new{also} complementary to those.
\new{Other recent approaches allow to identify the number of phases manually by adjusting a threshold \cite{AH23}, and consequently extract the period visually. In contrast, \approach not only automatically extracts the period and provides a metric for confidence but also performs this during an application's runtime in the online mode. Still, a combination of both approaches might be very valuable for the HPC community. }

In the field of performance analysis, Casas et al.~\cite{doi:10.1177/1094342009360039}
proposed to construct signals of metrics (e.g., number of active processes, amount of communicated data, etc.),
and then to apply discrete wavelet transform to keep the highest-frequency portions and autocorrelation to find the frequency of the application's phases. We were inspired by the use of signal processing techniques, but our approach is different since we advocate a lightweight approach to concisely represent the period of the I/O behavior, whereas they aim at removing the effects of external noise to detect the phases that best represent the application. 
Yang et al.~\cite{9820616} introduced a metric to quantify the burstiness of I/O and apply it to traces from a production machine. They found that most traces presented a very high degree of burstiness; however, their metric is a measure of ``unevenness'', not of periodicity, which is our focus.
Qiao et al~\cite{9355326} used DFT on a signal of write performance over function calls. They used it to search 
for the period of other concurrent  applications (and use that to predict future interference). 
We argue for a scenario where this information can be easily obtained for all applications and shared so that smart decisions can be made throughout the system.

%% file: 06_conclusion.tex
\section{Conclusion} 
\label{sec:conclusion}
\label{sec:beyond}
This paper presents \approach, an approach  
for characterizing and predicting the temporal I/O behavior of an application with a
simple metric, namely its period, obtained using different frequency techniques. 
We provided several extensions to adapt to the unsteady nature of I/O and further describe the behavior. 
Our evaluation demonstrates the low overhead of \approach, its suitability for real large-scale examples, and its robustness with a mean error below 11\%. Combined with the I/O scheduler Set-10, \approach allowed increasing system utilization by \new{26\%} and decreasing I/O slowdown by \new{56\%}. 
However, I/O scheduling is only one possible application of \approach. Its predictions could also be helpful in other contexts, such as burst buffer management \new{ (e.g., flushing before the buffer is full to overcome storage space restrictions)}. Moreover, the post-mortem analysis could be used for I/O-aware batch scheduling. 

Despite our focus on whole applications, there are use cases (e.g., cache management) which require knowing the behavior of individual processes. Even in such cases, our approach is equally suitable.
Furthermore, although we focused on I/O in this paper, 
our technique can be repurposed for other use cases (e.g., finding the period of scheduling points) by simply changing the input data. 
%
Finally, our discussion on the selection of the sampling frequency (Section~\ref{sec:param}) assumes we are interested in
\emph{any} frequency the application's I/O behavior exhibits. In some cases,
such as I/O scheduling, for example, we may \emph{not} be interested in high frequencies because we cannot respond fast enough, so the sampling frequency ($f_s$) could act as a filter. 
%
Future work will focus on exploring online $f_s$ adaptation. Moreover, our approach rests on DFT, which has a high-frequency resolution but no time resolution. 
We plan to explore 
merging the result 
with the wavelet transform~\cite{388960} for a more comprehensive characterization, to prepare for cases where we need both.

%% file: 07_acknowledgement.tex
\section*{Acknowledgment}
\label{sec:acknowledgements}

The authors would like to thank Jean-Baptiste Besnard (ParaTools) for his support and enthusiasm for this work. 
The authors also thank Clément Barthélemy and Luan Teylo for their help in setting up the Set-10 experiments.

\section*{Appendix: Artifacts Reproducibility}
As mentioned at the beginning of \cref{sec:ftio}, FTIO and TMIO are publicly available on GitHub. 
The FTIO repository\footnote{\url{https://github.com/tuda-parallel/FTIO/tree/main/artifacts/ipdps24}} provides descriptions on how to reproduce the results and experiments from this paper.
Additionally, the data sets from these experiments are publicly available~\cite{tarraf_2024_10670270}.